\documentclass[%
 reprint,
 amsmath,amssymb,
 nofootinbib,
 aps,
]{revtex4-2}

\usepackage{hyperref}
\hypersetup{
    colorlinks=true,       
    linkcolor=blue,          
    citecolor=red,        
    urlcolor=blue           
}

\usepackage{graphicx}
\usepackage{dcolumn}
\usepackage{bm}
\usepackage{amsmath}
\usepackage{color}

\usepackage{ulem}
\newcommand{\CB}{\color{blue}}

\newcommand{\CR}{\color{red}}

\begin{document}

\preprint{APS/123-QED}

\title{Relativistic non-resistive viscous magnetohydrodynamics from the kinetic theory:~a relaxation time approach}

\author{Ankit Kumar Panda}
\email{ankitkumar.panda@niser.ac.in}
\author{Ashutosh Dash}%
 \email{ashutosh.dash@niser.ac.in}
 \author{Rajesh Biswas}%
 \email{rajeshphysics143@gmail.com}
 \author{Victor Roy}%
 \email{victor@niser.ac.in}
\affiliation{%
 National Institute of Science Education and Research, Bhubaneswar, HBNI, Jatni, 752050, India
}%

\date{\today}

\begin{abstract}
We derive the relativistic non-resistive, viscous second-order magnetohydrodynamic
equations for the dissipative quantities using the relaxation time approximation.
The Boltzmann equation is solved for a system of particles and antiparticles using
Chapman-Enskog like gradient expansion of the single-particle distribution function
truncated at second order. In the first order, the transport coefficients are independent
of the magnetic field. In the second-order, new transport coefficients that couple magnetic
field and the dissipative quantities appear which are different from those obtained in the
14-moment approximation~\cite{Denicol:2018rbw} in the presence of a magnetic field.
However, in the limit of the weak magnetic field, the form of these equations are identical
to the 14-moment approximation albeit with a different values of these coefficients.
We also derive the anisotropic transport coefficients in the Navier-Stokes limit.
\end{abstract}

\maketitle

\section{\label{sec:intro}Introduction}
The ubiquitous magnetic field seems to have played a great role in shaping and working of
our present-day universe. We see magnetic fields at the very largest scales in the universe.
They are usually very weak, no more than a million times weaker than Earth's magnetic field,
but they exist throughout the known universe. Although weak, often the magnetic fields
are called the sturdy unsung workhorses of astrophysics and cosmology.
On the other hand, one of the strongest steady-state magnetic field in the known universe 
can be found on the surface of a type of neutron star called Magnetars.
Surprisingly, the strongest transient magnetic fields in the universe are manmade
and can be found on earth during the initial stages of heavy-ion collisions at Relativistic
Heavy Ion Collider~(RHIC) near Brookhaven, New York and at Large Hadron Collider~(LHC) near
Geneva, Switzerland.  These strongest magnetic fields are produced by fast-moving charged protons~(usually having Lorentz factor $\gamma \sim 100$ or larger) inside the colliding
nuclei of heavy ions~(Pb or Au). The strength of the magnetic field  produced in such a collider
experiments for example for  a typical peripheral Au+Au collisions at $\sqrt{s_{NN}}=200$ GeV may
reach as high as~($\sim 10^{18}\mbox{-}10^{19}$ Gauss)~\cite{Bzdak:2011yy,Deng:2012pc,Tuchin:2013ie,Roy:2015coa,Li:2016tel}, this is almost
three to four orders of magnitude larger than those found on Magnetars.
The magnitude of the produced magnetic field is expected to grow linearly with
the center-of-mass energy but the lifetime of these strong fields reduces for higher energy collisions.
The heavy-ion collisions also produce a new form of very hot and dense matter known as quark-gluon plasma~(QGP).
The success of relativistic hydrodynamics in describing the space-time evolution
of the QGP created in high-energy heavy-ion collisions~\cite{Chaudhuri:2012yt,Heinz:2013th,Gale:2013da,Romatschke:2017ejr} and the existence of very large magnetic
fields in these collisions indicates that one should take into account the proper interaction of magnetic
fields with QGP. Especially since the QGP and the subsequent hadronic phase are known to be
electrically conducting~\cite{Gupta:2003zh,Aarts:2014nba,Amato:2013naa}.

Relativistic magnetohydrodynamics~(RMHD) is one of the self-consistent macroscopic frameworks
that describes the evolution of mutually interacting charged fluid and electromagnetic fields.
In several recent works, the effect of the electromagnetic fields on the QGP fluid in the context of
special relativistic systems have been studied~\cite{Huang:2009ue,Huang:2015oca,Greif:2017irh,Roy:2017yvg,Gursoy:2014aka,Huang:2015oca,Inghirami:2016iru,Huang:2011dc}. Almost all of them involves
numerical solutions of RMHD equations, the analytic solutions for some simplified cases are presented in~\cite{Roy:2015kma,Pu:2016ayh,Siddique:2019gqh,Pu:2016bxy,Pu:2016ayh,Shokri:2018qcu,Shokri:2017xxn,Moghaddam:2017myy}. The transport coefficients  such as the shear, bulk viscosity etc. are taken as input to the RMHD simulation,
but they are determined from an underlying microscopic theory~\cite{Arnold:2000dr,Arnold:2003zc,Li:2017tgi,Ghosh:2018cxb,Cao:2018ews,Kurian:2018qwb,Singh:2017nfa,Bhadury:2020ivo,Chakrabarty:1986xx}. It is a well known fact that a straightforward extension of non-relativistic viscous fluid formulations
(a.k.a. Navier-Stokes equation) to the relativistic regime (without magnetic field)~\cite{Eckart:1940te,LL} leads to
unacceptable acuasal and linearly unstable behaviour~\cite{Hiscock:1983zz,Hiscock:1985zz,Hiscock:1987zz}.
These issues were later addressed and resolved by Israel and Stewart~(IS)
who developed a causal and stable second-order formalism~\cite{Israel:1976tn,Israel:1979wp}. The order of the theory
is determined by the presence of different order terms in the gradient expansion of the
hydrodynamic quantities such as fluid four-velocity $u^{\mu}$, temperature $T$, etc. in the energy-momentum tensor.
Although IS resolves the major problem, the theory is known to be
causal and stable in a restricted manner~\cite{Pu:2009fj,Denicol:2008rj,Denicol:2008ha,Floerchinger:2017cii}.
Recently, there are some new developments in the formulation of first-order casual and theories
which is claimed causal and stable in a restricted sense~\cite{Van:2011yn,Kovtun:2019hdm,Bemfica:2019knx,Das:2020fnr,Taghinavaz:2020axp,Hoult:2020eho}. However, we note that in the newly developed
 theory the existence of a relaxation time scale (usually found in the second-order theories)
 in the definition of non-equilibrium hydrodynamics variables  needs further investigation.
Although initially developed as a phenomenological theory, the IS theory was later
derived from the underlying kinetic theory using Grad's moment method. One of the limitations
of the moment method is the absence of a smallness parameter such as Knudsen number~(Kn)
which otherwise would have helped to systematically improve the result by keeping higher-order terms.
Later, more concrete and updated form of the IS equations (without electromagnetic fields) were derived from
the kinetic theory~\cite{Betz:2008me,Denicol:2010xn,Muronga:2006zx,Denicol:2012cn,Denicol:2012es,Betz:2009zz,York:2008rr,Jaiswal:2012qm,Jaiswal:2013npa,Florkowski:2010cf}.

It was only recently that the second-order causal magnetohydrodynamics equations were derived for
non-resistive~\cite{Denicol:2018rbw} and resistive case~\cite{Denicol:2019iyh}
for a single component system of spinless particles (no antiparticle) using a $14$-moment approximation.
In the current work we consider contribution from both particles and antiparticles,
henceforth whenever we compare our results with \cite{Denicol:2018rbw}  we report the result for particles only.
In Ref.~\cite{Biswas:2020rps} we showed that this new theory of second-order relativistic MHD
is causal and stable under linear perturbation.
In this paper, we derive the RMHD equations for the non-resistive case using Chapman-Enskog
expansion of the single-particle distribution function within relaxation time approximation~(RTA).
Here we consider both particles and antiparticles while calculating the 
relaxation equations for the dissipative quantities.
Due to the presence of smallness parameter $\mathrm{Kn}$ in the RTA formalism we have the freedom
to construct magnetohydrodynamics equations order by order and calculate corresponding transport
coefficients. It is necessary to use a causal theory of magnetohydrodynamics to study other
important phenomena associated with strong magnetic fields. For example,
the coexistence of the strongest magnetic field in the universe and the hot dense
medium of quarks and gluons also opens up possibilities to experimentally verify some of the
fundamental issues of Quantum Chromo-Dynamics~(QCD). One such fascinating
phenomena is ``chiral magnetic effect''~(CME) where an induced charge current is supposed to
be produced parallel to the magnetic field in a chiral imbalance system~\cite{Kharzeev:2004ey,Fukushima:2008xe}.
Some other important phenomenon associated with strong magnetic fields are
chiral separation effect~\cite{Son:2004tq}, chiral hall effect~\cite{Pu:2014fva},
chiral vortical effect~\cite{Banerjee:2008th} etc. A chiral kinetic theory framework
is currently under development to further explore these important phenomena~\cite{Stephanov:2012ki,Chen:2014cla,Hidaka:2016yjf}.

The manuscript is organized as follows:~In Sec.~\ref{sec:sec2} we give a textbook-like introduction to the energy-momentum
tensor for the electromagnetic fields and the fluid-matter, we also discuss the kinetic theory definition of various hydrodynamical variables in the same section. In Sec.~\ref{sec:sec3} we present the first and second order
magneto-hydrodynamic equations of motion for the non-resistive dissipative fluid. We conclude this work
in Sec.~\ref{sec:conclusion}.
Throughout the paper we use the natural units, $\hbar=c=k_{B}=\epsilon_{0}=\mu_{0}=1$ and the metric tensor
in flat space-time is $g^{\mu\nu}=$diag$\left(+1,-1,-1,-1\right)$. The time-like fluid four velocity $u^{\mu}$ satisfy
$u_{\mu}u^{\mu}=1$. Also, we use the following decomposition for the partial derivative: $\partial_{\mu}\equiv u_{\mu}u_{\nu}\partial^{\nu}+(g_{\mu\nu}-u_{\mu}u_{\nu})\partial^{\nu}=u_{\mu}D+\nabla_{\mu}$. The
 $\nabla^{\alpha}u^{\beta}$ is decomposed as:
\begin{equation}
\label{eq:nablaanduexpansion}
  \nabla^{\alpha}u^{\beta}=\omega^{\alpha\beta}+\sigma^{\alpha\beta}+\frac{1}{3}\theta\Delta^{\alpha\beta},
\end{equation}
where $\omega^{\alpha\beta}=(\nabla^{\alpha}u^{\beta}-\nabla^{\beta}u^{\alpha})/2$ is the
anti-symmetric vorticity tensor, $\sigma^{\alpha\beta}\equiv \nabla^{\langle\mu} u^{\nu\rangle}
=\frac{1}{2}\left(\nabla^{\mu} u^{\nu}+\nabla^{\nu} u^{\mu}\right)-\frac{1}{3}
\theta \Delta^{\mu \nu}$ is the symmetric-traceless tensor and
$\theta \equiv \partial_{\mu} u^{\mu}$ is the expansion scalar.
The fourth-rank projection tensor is defined as
$\Delta^{\mu\nu}_{\alpha\beta}=\frac{1}{2}\left(\Delta^{\mu}_{\alpha}\Delta^{\nu}_{\beta}+\Delta^{\mu}_{\beta}\Delta^{\nu}_{\alpha}\right)-\frac{1}{3}\Delta^{\mu\nu}\Delta_{\alpha\beta}$.

\section{\label{sec:sec2}RELATIVISTIC MAGNETOHYDRODYNAMICS}

\subsection{Equation of motion of the electromagnetic field}
Here we start by giving some text-book like introduction to the relativistically
covariant formulation of electrodynamics. Without any loss of generality the second rank
antisymmetric electromagnetic field tensor $F^{\mu\nu}$ can be defined
in terms of the electric $E^{\mu}$, magnetic field $B^{\mu}$ four-vectors~(defined later) and
four-velocity $u^{\mu}$ as in Refs.~\cite{Lichnerowicz,Anile,Kip}:
\begin{equation}
\label{eq:emtensor}
  F^{\mu\nu}=E^{\mu}u^{\nu}-E^{\nu}u^{\mu}+\epsilon^{\mu\nu\alpha\beta}u_{\alpha}B_{\beta},
\end{equation}
its dual counter-part is given by:
\begin{equation}
\label{eq:dual}
  \tilde{F}^{\mu\nu}=B^{\mu}u^{\nu}-B^{\nu}u^{\mu}-\epsilon^{\mu\nu\alpha\beta}u_{\alpha}E_{\beta},
\end{equation}
where $E^{\mu}=F^{\mu\nu}u_{\nu}$ and
$B^{\mu}=\tilde{F}^{\mu\nu} u_{\nu}=\frac{1}{2} \epsilon^{\mu \nu \alpha \beta} u_{\nu} F_{\alpha \beta} $.
Also, using the anti-symmetric property of the $F^{\mu\nu}$ it is easy to see that both $E^{\mu}$ and $B^{\mu}$
are orthogonal to $u^{\mu}$ i.e., $E^{\mu}u_{\mu}=B^{\mu}u_{\mu}=0$. Futhermore notice that in the
rest frame $u^{\mu}=(1,\bf 0)$ we have $E^{\mu}:=(0,\bf E)$, and $B^{\mu}:=(0,\bf B)$, where $\bf E,B$
corresponds to the electric and magnetic field three vectors with $\mathrm{E}^{i}:=F^{i 0}$ and
$\mathrm{B}^{i}:=-\frac{1}{2} \epsilon^{i j k} F_{j k}$.

We can write the  Maxwell's equations in a covariant form as:
\begin{eqnarray}
\label{eq:maxwelleqn1}
  \partial_{\mu}F^{\mu\nu} &=& J^{\nu}, \\
\label{eq:maxwelleqn2}
  \partial_{\mu}\tilde{F}^{\mu\nu} &=& 0,
\end{eqnarray}
where $J^{\nu}$ is the electric charge four-current which acts as the source of electromagnetic field.
It can be tensor decomposed in a fluid with four velocity $u^{\mu}$ in the following manner:
\begin{equation}
\label{eq:chargecurrent}
 J^{\mu}=j^{\mu}+d^{\mu},
\end{equation}
where $j^{\mu}$ is the conduction current and
$d^{\mu}=\Delta^{\mu}_{\nu}J^{\nu}$ is the charge diffusion current
with $n_{q}=u_{\mu}J^{\mu} $ the proper net charge density.
If we assume a linear constitutive relation between $j^{\mu}$ and $E^{\mu}$~(Ohm's law) then
$ j^{\mu}=\sigma^{\mu\nu}E_{\nu}$ where $\sigma^{\mu\nu}$ is the conductivity tensor.
Also note that by construction $u_{\mu}j^{\mu}=0$ which imply that the conduction current
exists even for the vanishing net charge.
The solution of Eqs.~\eqref{eq:maxwelleqn1},~\eqref{eq:maxwelleqn2} along with a given
$J^{\mu}$ in Eq.~\eqref{eq:chargecurrent} completely specify the electro-magnetic field evolution.
$J^{\mu}$ acts as a coupling between the fluid and the fields because it contains the fluid
informations such as fluid conductivity $\sigma^{\mu\nu}$, net charge density $n_{q}$ etc., and
act as a source in the Maxwell's equations. Incidentally, for a single component gas as
considered here the net charge is equivalent to net number density and the following relation holds
$n_{q}=q n_{f}$, where $n_{f}$ corresponds to net number density.

We assume here that the fluid under consideration does not possess polarisation or magnetisation
and thus the electromagnetic field stress-energy tensor can be written as:
\begin{equation}\label{Temmunu}
  T^{\mu\nu}_{EM}=-F^{\mu\lambda}F^{\nu}_{\lambda}+\frac{1}{4}g^{\mu\nu}F^{\alpha\beta}F_{\alpha\beta}.
\end{equation}
Now taking the partial derivative of the field stress-energy tensor we get the equation of motion to be:
\begin{equation}\label{partialTmunu}
  \partial_{\mu}T^{\mu\nu}_{EM}=-F^{\nu\lambda}J_{\lambda}.
\end{equation}
Up until now, we consider the system under consideration has the charge density
of only fluid and there are no external sources of electromagnetic fields i.e., $J^{\mu}=J^{\mu}_f$.
However, in presence of an external source current $J^{\mu}_{ext}$~(for example, the spectator
protons in heavy-ion collisions act like an external source for the electromagnetic fields in the QGP) the total current is
a combination of conduction and external current densities:
\begin{equation}\label{totalchargedensity}
  J^{\mu}=J^{\mu}_f +J^{\mu}_{ext}.
\end{equation}
In this case, the external current density acts as a source term in the energy-momentum conservation
equation~(discusses later in detail).
In this work, we consider an ideal MHD limit which corresponds to very large magnetic Reynolds number
$R_{m} \gg 1$. The magnetic Reynolds number is given as, $R_{m} = L U \sigma \mu$, where $L$ is the
characteristic length or time scale of the QGP, $U$ is the characteristic velocity of the flow
and $\mu$ is the magnetic permeability of QGP. The large $R_m$ limit can be attributed to a very
large/infinite electrical conductivity. But the induced charge density due to the electromagnetic
field $J^{\mu}_{ind}=\sigma E^{\mu}$ ~(here $\sigma$
is the isotropic electrical conductivity i.e., $\sigma^{\mu\nu}=\sigma g^{\mu\nu}$)
has to be finite, so to maintain that  $E^{\mu}\rightarrow 0$ for this case.
This brings our electromagnetic tensor $F^{\mu\nu}$ to the following form:
\begin{equation}
\label{eq:modemtensor}
  F^{\mu\nu}\rightarrow B^{\mu\nu}=\epsilon^{\mu\nu\alpha\beta}u_{\alpha}B_{\beta}.
\end{equation}
Using  Eqs.~\eqref{totalchargedensity} and~\eqref{eq:modemtensor} in
the Maxwell's equations Eq.~\eqref{eq:maxwelleqn1} we get:
\begin{equation}
\label{eq:EoMmaxmodified}
  \epsilon^{\mu\nu\alpha\beta}\left(u_{\alpha}\partial_{\mu}B_{\beta}+B_{\beta}\partial_{\mu}u_{\alpha}\right)=J^{\nu}_{f} +J^{\nu}_{ext}.
\end{equation}
Now writing the energy-momentum tensor for the electromagnetic case by using Eqs.~\eqref{Temmunu} and~\eqref{eq:modemtensor} we get:
\begin{equation}\label{emtensorforB}
  T^{\mu\nu}_{EM}\rightarrow T^{\mu\nu}_B=\frac{B^2}{2} \left(u^{\mu}u^{\nu}-\Delta^{\mu\nu}-2b^{\mu}b^{\nu}\right),
\end{equation}
where $B^{\mu}B_{\mu}=-B^{2}$ and $b^{\mu}=\frac{B^{\mu}}{B}$ with the property
 $b^{\mu}u_{\mu}=0$ and $b^{\mu}b_{\mu}=-1$. Furthermore from Eq.~\eqref{eq:modemtensor}
 one can show that $B^{\mu\nu}B_{\mu\nu}=2B^{2}$
 so we can introduce another anti-symmetric tensor defined as:
\begin{equation}\label{bmunu}
  b^{\mu\nu}=-\frac{B^{\mu\nu}}{B},
\end{equation}
with the following properties: $b^{\mu\nu}u_{\nu}=b^{\mu\nu}b_{\nu}=0$ and $b^{\mu\nu}b_{\mu\nu}=2$.
\subsection{Kinetic theory and hydrodynamics}
In this section, we define a few hydrodynamical variables from the kinetic theory.
We start with the equilibrium distribution function for particles given by $f_0$ and is defined as:
\begin{equation}\label{distfunc}
  f_0=\frac{1}{e^{\beta(u \cdot p) - \alpha}+r},
\end{equation}
where $r=+1$ for fermions and $r=-1$ for bosons and $r=0$ for Boltzmann gas.
Here $\beta=\frac{1}{T}$ is the inverse temperature, $u^{\mu}$ is the four-velocity, $p^{\mu}$ is
the four-momentum and $\alpha=\frac{\mu}{T}$  is the ratio of chemical potential to temperature
with $\mu$ being the chemical potential. For antiparticles
$\alpha \rightarrow -\alpha$ and $f_{0}\rightarrow \bar{f_{0}}$.

For a dissipative fluid~(as is considered in this work) one needs to fix the definition of $u^{\mu}$.
One popular choice is the Landau-Lifshitz frame where the $u^{\mu}$ is defined such that the heat
flux vanishes in the local rest frame of the fluid; in that case, the net four current $N^{\mu}$
and the energy-momentum  tensor $T_f^{\mu\nu}$ can be decomposed in  terms of $u^{\mu}$,
the projector operator $\Delta^{\mu\nu}$, and dissipative fluxes diffusion current $V_{f}^{\mu}$,
the shear $\pi^{\mu\nu}$ and bulk stress $\Pi$ in the following way:
\begin{eqnarray}
   \label{Nmu}
   N^{\mu} &=& n_{f} u^{\mu}+V_{f}^{\mu}, \\
   \label{Tfmunu}
   T_f^{\mu\nu} &=& \epsilon u^{\mu}u^{\nu}-(P+\Pi)\Delta^{\mu\nu}+\pi^{\mu\nu},
\end{eqnarray}
where $n_f$ is the net number density, $\epsilon$ is the energy density, $P$ is the isotropic pressure of fluid.
According to kinetic theory framework the energy-momentum tensor and the particle four current
of a fluid can be defined in terms of moments of the single particle distribution function $f$ in
 the following way:
\begin{eqnarray}{\label{integraldeffluid}}
   T^{\mu\nu}_f &=& \int dp p^{\mu}p^{\nu}\left(f+\bar{f}\right), \\
   N^{\mu} &=& \int dp  p^{\mu}\left(f -\bar{f}\right),
\end{eqnarray}
where $dp = g d^3 \mathbf{p}/[(2\pi)^3 p^0]$ with $p^0 =\sqrt{\mathbf{p}^2 +m^2}$, $m$
 being the mass, $g$ is the degeneracy factor.  For an out of equilibrium system the dustribution
 function can be decomposed into equilibrium $f_{0}$ and a correction to it $\delta f$ as
 $f=f_{0}+\delta f $ (for antiparticle $\delta f \rightarrow \delta \bar{f}$).
 The explicit form of the $\delta f $ is obtained from the Boltzmann equation and it depends on the
 scheme used. For example the $\delta f$ for the case of relativistic ideal gas in magnetic field
 in terms of Grad's fourteen moment method was derived in Ref.~\cite{Denicol:2018rbw}.
With the above definition of $N^{\mu}$ and $T^{\mu\nu}$ all other thermodynamic variables
can be defined as:
\begin{eqnarray}
\label{eq:energyKin}
  &&\epsilon \equiv u_{\mu}u_{\nu}T^{\mu\nu}=u_{\mu}u_{\nu}\int_{}^{}dp p^{\mu}p^{\nu} \left(f_0 +\bar{f_0}\right), \\
  \label{eq:numberKin}
  &&n_f \equiv u_{\mu}N^{\mu} = u_{\mu}\int_{}^{}dp p^{\mu}\left(f_0 -\bar{f_0}\right), \\
  \label{eq:pressKin}
  &&P \equiv -\frac{\Delta_{\mu\nu}}{3}T^{\mu\nu} = -\frac{\Delta_{\mu\nu}}{3}\int_{}^{}dp p^{\mu}p^{\nu}\left(f_0 +\bar{f_0}\right), \\
  &&V_f^{\mu} \equiv  \Delta^{\mu}_{\nu} N^{\nu}=\Delta^{\mu}_{\nu}\int_{}^{}dp p^{\nu}\left(\delta f -\delta \bar{f}\right), \\
  &&\Pi \equiv-\frac{\Delta_{\mu\nu}}{3}\delta T^{\mu\nu}= -\frac{\Delta_{\mu\nu}}{3}\int_{}^{}dp p^{\mu}p^{\nu}\left(\delta f +\delta \bar{f}\right), \\
  &&\pi^{\mu\nu} \equiv \Delta^{\mu\nu}_{\alpha\beta}\delta T^{\mu\nu} = \Delta^{\mu\nu}_{\alpha\beta}\int dp p^{\alpha}p^{\beta}\left(\delta f+\delta \bar{f} \right),
\end{eqnarray}
where $\delta T^{\mu\nu}=-\Pi\Delta^{\mu\nu}+\pi^{\mu\nu}$.
For a single component fluid the net current $N^{\mu}$ and charged current are related as:
\begin{equation}\label{rlnchargecurrentandparticlecurrent}
  J_f^{\mu}=q N^{\mu},
\end{equation}
where $q$ is the magnitude of electric charge.

For later use, we express the integrals in  Eq.~\eqref{eq:energyKin} to Eq.~\eqref{eq:pressKin}
 in terms of thermodynamics integrals $I_{nq}^{(m)\pm}$ (defined in Appendix~\ref{appendix1}) as:
\begin{eqnarray}
  \epsilon &=& I_{20}^{(0)+}, \\
  n_f &=& I_{10}^{(0)-}, \\
  P &=& -I_{21}^{(0)+}.
\end{eqnarray}
here $\pm$ corresponds to the addition or subtraction of $\bar{f}$.
\subsection{Equation of motion of magnetohydrodynamics}
\subsubsection{Conservation of energy and momentum of fluid and electromagnetic field}
In a simple fluid (for zero magnetic field) the energy-momentum tensor and the particle currents
are conserved separately according to the following conservation law:
\begin{eqnarray}
  \partial_{\mu}N^{\mu} &=& 0, \\
  \partial_{\mu}T_f^{\mu\nu} &=& 0.
\end{eqnarray}
Now let us consider a fluid interacting with the electro-magentic field
and let $T^{\mu\nu}$ be the total energy-momentum tensor (field+fluid).
$T^{\mu\nu}$ can be written as a sum of
energy-momentum tensor of the fluid and the electromagnetic field as:
\begin{equation}
\label{eq:totem}
T^{\mu\nu}=T^{\mu\nu}_f+T^{\mu\nu}_{EM}.
\end{equation}
In general, the total energy-momentum tensor contains additional terms in Ref.~\cite{Anile}
which cannot be unambiguously attributed to either fluid or field but in case of constant
susceptibilities and vanishing $E^{\mu}$ these terms vanishes and Eq.~\eqref{eq:totem}
is a good approximation. Note that due to the conservation of electric
charges, the charge current of the fluid is individually conserved:
\begin{equation}\label{concharge}
  \partial_{\mu}J^{\mu}_f=0.
\end{equation}
If we have an external charge current, it will act as a source in the energy momentum conservation
equation which in this case takes the following form:
\begin{equation}\label{partialtot}
  \partial_{\mu}T^{\mu\nu}=-F^{\nu\lambda}J_{ext,\lambda}.
\end{equation}
 The conservation equation for electromagnetic field Eq.~\eqref{partialTmunu} with external source
 takes the following form:
\begin{equation}\label{partialem}
  \partial_{\mu}T^{\mu\nu}_{EM}= -F^{\nu\lambda}\left(J_{f,\lambda}+J_{ext,\lambda}\right).
\end{equation}
Using Eq.~\eqref{eq:totem} and Eqs.~\eqref{partialtot},~\eqref{partialem} we get:
\begin{equation}\label{partialfluid}
  \partial_{\mu}T^{\mu\nu}_f= F^{\nu\lambda}J_{f,\lambda}.
\end{equation}
Usually, the total energy-momentum tensor of an isolated system remains conserved but in case of
the presence of an external source (here external charge current) the conservation is satisfied only when a proper
source term is taken into account.  As we can see that in this case, the fluid evolution depends on the
fluid charge current  through Eq.~\eqref{partialfluid}.

 It is convenient to express the conservation equations in an alternative form by taking
projection along and perpendicular to fluid four velocity.
The parallel projection of Eq.~\eqref{partialem} and Eq.~\eqref{partialfluid} gives:
\begin{eqnarray}
  \label{eq:paralEM}
  u_{\nu}\partial_{\mu}T^{\mu\nu}_{EM} &=& 0, \\
  \label{eq:paralFl}
  u_{\nu}\partial_{\mu}T^{\mu\nu}_f &=& 0,
\end{eqnarray}
it implies that the energy density of the fluid and the
field are unaffected by the charge currents/magnetic field.
The perpendicular projection of Eq.~\eqref{partialem} and Eq.~\eqref{partialfluid} using Eq.~\eqref{totalchargedensity} gives
\begin{eqnarray}
  \label{eq:perpEM}
  \Delta^{\alpha}_{\nu}\partial_{\mu}T^{\mu\nu}_{EM}&=& Bb^{\alpha\lambda}\left(J_{f,\lambda}+J_{ext,\lambda}\right), \\
  \label{eq:perpFL}
  \Delta^{\alpha}_{\nu}\partial_{\mu}T^{\mu\nu}_f &=& -Bb^{\alpha\lambda}J_{f,\lambda}.
\end{eqnarray}
This shows that unlike energy density, the momentum density of the fluid depends
on the diffusion current/magnetic field and the momentum density of the field
also depends on the external current, along with the fluid diffusion current.

\subsubsection{Ideal and dissipative non-resistive magnetohydrodynamics}
In case the fluid is ideal, the total energy momentum tensor takes the form:
 \begin{equation}\label{idealmhd}
   T^{\mu\nu}_{(0)}=\left(\epsilon+\frac{B^2}{2}\right)u^{\mu}u^{\nu}-\left(P+\frac{B^2}{2}\right)\Delta^{\mu\nu}-B^2b^{\mu}b^{\nu}.
 \end{equation}

If the fluid is dissipative with finite shear and bulk viscosity, the energy-momentum tensor in that case becomes:
\begin{equation}
{\label{eq:fullEM}}
T^{\mu\nu} = \left(\epsilon+\frac{B^2}{2}\right)u^{\mu}u^{\nu}-\left(P+\Pi+\frac{B^2}{2}\right)\Delta^{\mu\nu}-B^{\mu}B^{\nu}+\pi^{\mu\nu}.
\end{equation}
The system of equations is closed with the constitutive relation of charged-current
  $J^{\mu}_f=\boldsymbol n_{f}u^{\mu}+d^{\mu}_f$ and with an Equation of State (EoS) relating thermodynamic
  pressure to energy and number density $p=p(\epsilon,n_{f})$.
Now using Eq.~\eqref{concharge}, Eq.~\eqref{partialfluid} along with Eq.~\eqref{Nmu},
Eq.~\eqref{Tfmunu} and using the thermodynamic integrals given in
Eq.~\eqref{eq:J} and Eq.~\eqref{relJI} we get the evolution equations for $\dot{\alpha}$, $\dot{\beta}$
and $\dot{u}^{\mu}$ which are of the following forms:
\begin{widetext}
\begin{eqnarray}
{\label{eq:alphadot}}
&&\dot{\alpha} = \frac{1}{D_{20}}\left[-J_{30}^{(0)+}(n_f \theta +\partial_{\mu}V_f^{\mu})+J_{20}^{(0)-}\left\{\ \left(hn_{f}+\Pi\right)\theta -\pi^{\mu\nu}\sigma_{\mu\nu}\right\}\ \right],\\
{\label{eq:betadot}}
&&\dot{\beta} = \frac{1}{D_{20}}\left[-J_{20}^{(0)-}(n_f \theta +\partial_{\mu}V_f^{\mu})+J_{10}^{(0)+}\left\{\ \left(hn_{f}+\Pi\right)\theta-\pi^{\mu\nu}\sigma_{\mu\nu}\right\}\ \right],\\
{\label{eq:udot}}
&&\dot{u}^{\mu} = \frac{1}{\left(1+\Pi\right) hn_{f}}\left[\frac{n_f}{\beta}\left(\nabla^{\mu}\alpha-h \nabla^{\mu}\beta \right)-\Delta^{\mu}_{\nu}\partial_{\gamma}\pi^{\gamma \nu} +\nabla^{\mu}\Pi-q Bb^{\mu\nu}V_{f,\nu}\right],
\end{eqnarray}
\end{widetext}
where $D_{20}=J_{30}^{(0)+}J_{10}^{(0)+}-J_{20}^{(0)-}J_{20}^{(0)-}$ , $h=\frac{\epsilon+P}{n_{f}}$.
\section{\label{sec:sec3}Formalism and results}
\subsection{Boltzmann Equation}{\label{sec:l31}}
The relativistic Boltzmann equation (RBE)  in the presence of a non-zero force $\mathcal{F}^{\nu}$ is given by:
\begin{equation}\label{RBE}
  p^{\mu}\partial_{\mu}f +\mathcal{F}^{\nu}\frac{\partial}{\partial p^{\nu}}f= C[f],
\end{equation}
where  $f({\bf x,p},t)$ is the one particle distribution function characterising
the phase space density of the particles,  $C[f]$
is the collision kernel. The first term on the left hand side corresponds to
the free streaming of the phase space density, the second term corresponds
to the effect of forces on the movement of
these phase space points, the collision kernel on the right hand side contributes
to the change in phase space density in an elemental volume due to the change
in momentum and positions of the colliding particles from their straight line free
streaming trajectories.
In the ideal MHD limit the electric field vanishes in the local rest frame of the fluid,
hence the only contribution to the force term in the RBE is due to the magnetic field
which is $\mathcal{F}^{\nu}:=qF^{\nu\alpha}p_{\alpha}$ for particles 
where $q$ is the electric charge of the particles,
and $F^{\mu\nu}=-Bb^{\mu\nu}$ (Eq.~\eqref{eq:modemtensor}).

The collision kernel is a non-linear term containing the product of the single particle
distribution function and it creates difficulty for solving the Boltzmann equation in
a closed form. Much simplification can be made if we assume the collision kernel of the
form relaxation time approximation given by Bhatnagar-Gross-Krook (BGK) for non-relativistic
systems and by Anderson-Witting in Ref.~\cite{Anderson} for the relativistic systems of the following form
 $C[f]=-\frac{u \cdot p}{\tau_c}\delta f$ where $\tau_{c}$ is the relaxation time or the
time taken by the particles away from equilibrium to come to the equilibrium state
 and $\delta{f} = f-f_{0}$ denotes the deviation from the equilibrium distribution $f_{0}$.
 In other words, in relativistic BGK-RTA approach the full momentum dependence of relevant scattering rates are
 approximately characterized by a single relaxation time $\tau_c$. Now putting this collision kernel
 into the Eq.~\eqref{RBE} we get the RBE for the particles:
\begin{equation}
\label{modrbe}
  p^{\mu}\partial_{\mu}f \pm qF^{\sigma\nu}p_{\nu}\frac{\partial}{\partial p^{\sigma}}f= -\frac{u \cdot p}{\tau_c}\delta f.
\end{equation}
The corresponding equation for the antiparticles are obtained by replacing $q\rightarrow -q$ and $f\rightarrow \bar{f}$.
The above approximation of the collision kernel in the Boltzmann equation has its own limitation e.g.,
the relaxation time ($\tau_c$) here does not depend on momentum of the colliding particles as mentioned
above.The interaction between colliding particles are such that the mean free path is larger than the interaction length.
In other words,  $f({\bf x,p},t)$ is assumed to be  not varying much over a time interval larger than the
duration of collision but smaller than the time  between collisions. The same applies for the change in
 $f({\bf x,p},t)$ over distance of the order of interaction range.

\subsection{Expansion in gradients}
One can cast Eq.~\eqref{modrbe} to the well-known hydrodynamic gradient
expansion form in Ref.~\cite{Heller:2013fn}, given the system is close to equilibrium, i.e., the collision kernel is
almost vanishing, $C[f]\approx0$. In the absence of an electromagnetic field Eq.~\eqref{modrbe}
can be written in the following form:
\begin{equation}
 \left(\frac{\tau_c}{u \cdot p}p^\mu\partial_\mu +1\right)f=(\mathcal{D}+1)f=f_0
\end{equation}
where we have introduced the operator $\mathcal{D}\equiv \frac{\tau_c}{u \cdot p}p^\mu\partial_\mu$. Multiplying the inverse operator $(\mathcal{D}+1)^{-1}$ in the above equation and subsequently doing  a power series expansion gives
\begin{equation}
 f=\sum_{n=0}^{\infty}(\mathcal{-D})^n f_0=\sum_{n=0}^{\infty}\left(-\frac{\tau_c}{u \cdot p}p^\mu\partial_\mu\right)^n f_0.
\end{equation}
The above expansion is valid given that $\mathrm{Kn}=\tau_c\partial\ll 1$, which of course is also the relevant expansion parameter. If one identifies the typical gradient strength to  be proportional to the temperature, $\partial\sim T$, then the expansion parameter is $\tau_c T$ and the series expansion is valid for $\tau_cT\ll 1$.

However, in the presence of a magnetic field, the naive gradient expansion breaks since one introduces
a new scale into the problem which is proportional to the strength of the magnetic field. Defining the operator $\mathcal{D}_B\equiv\frac{\tau_c}{u \cdot p}\left(p^{\mu}\partial_{\mu}+qF^{\sigma\nu}p_{\nu}\frac{\partial}{\partial p^{\sigma}}\right)$ and doing a similar power series expansion, gives the following result
\begin{eqnarray}
\label{eq:expansion}
f &=& \sum_{n=0}^{\infty}(-\mathcal{D}_B)^n f_0 , \nonumber\\
&=&\sum_{n=0}^{\infty}{\left[-\frac{\tau_c}{u \cdot p}\left(p^{\mu}\partial_{\mu}+qF^{\sigma\nu}p_{\nu}\frac{\partial}{\partial p^{\sigma}}\right)\right]}^nf_0.
\end{eqnarray}
Along with previous assumption $\tau_c T\ll 1$, one has to also assume that $\tau_c/r_g\ll1$, where $r_g=k_{\bot}/qB$ is the gyroradius (Larmor radius) and $k_{\bot}$ is the component of the momentum perpendicular to the direction of the magnetic field. In the plasma, the typical transverse momentum of particle $k_{\bot}\sim T$ and thus one has also to satisfy the condition $\chi=qB\tau_c/T\ll 1$.
 In the following, $f$ is obtained by keeping the terms up to second order, i.e., $n=2$ in Eq.~\eqref{eq:expansion}, which yields

\begin{equation}
\label{eq:2ndorderF}
  f=f_0 + \delta f^{(1)} + \delta f^{(2)},
\end{equation}
where
\begin{equation}
 \delta f^{(1)}=-\frac{\tau_c}{u \cdot p}\left(p^{\mu}\partial_{\mu}+qF^{\sigma\nu}p_{\nu}\frac{\partial}{\partial p^{\sigma}}\right)f_0,
\end{equation}
and
\begin{eqnarray}
\delta f^{(2)}&=&\frac{\tau_c}{u \cdot p}\left(p^{\mu}\partial_{\mu}
 +qF^{\sigma\nu}p_{\nu}\frac{\partial}{\partial p^{\sigma}}\right) \times\nonumber\\
 &&\left[\frac{\tau_c}{u \cdot p}\left(p^{\alpha}\partial_{\alpha}+qF^{\rho\beta}p_{\beta}\frac{\partial}{\partial p^{\rho}}\right)f_0 \right].
\end{eqnarray}
The above expression can be simplified by using the relations $qBb^{\mu\nu}p_{\nu}\frac{\partial f_0}{\partial p^{\mu}}=0$ and $F^{\mu\nu}=-Bb^{\mu\nu}$, which gives
\begin{equation}
\label{eq:2ndorderFfinal}
  f=f_0 + \delta \tilde{f}^{(1)} + \delta \tilde{f}^{(2)},
\end{equation}
where
\begin{equation}
\label{eq:deltaf1}
  \delta \tilde{f}^{(1)} = -\frac{\tau_c}{u \cdot p}p^{\mu}\partial_{\mu}f_0,
\end{equation}
and
\begin{eqnarray}
\label{eq:deltaf2}
\nonumber
  \delta \tilde{f}^{(2)} &=& \frac{\tau_c}{u \cdot p}p^{\mu}\partial_{\mu}\left[\frac{\tau_c}{u \cdot p}p^{\alpha}\partial_{\alpha}f_0\right]\\
  & &-\frac{\tau_c}{u \cdot p}qBb^{\sigma\nu}p_{\nu}\frac{\partial}{\partial p^{\sigma}}\left[\frac{\tau_c}{u \cdot p}p^{\alpha}\partial_{\alpha}f_0\right].
\end{eqnarray}
Similarly for antiparticles $\delta \bar{f}$ is calculated by replacing $f_{0} \rightarrow \bar{f}_{0}$
and $q \rightarrow -q$.
It is important to note that although the magnetic field does not enter explicitly in the
first term of $\delta \tilde{f}^{(2)}$, moreover it does enter implicitly through the
acceleration term $\dot{u}^\mu$, i.e., Eq.~\eqref{eq:udot} while taking higher-order
moments of such terms.
\subsection{First order equations}
The term $\delta \tilde{f}^{(1)}$ neither depends explicitly nor implicitly on the magnetic field,
since in the first-order equations we keep terms till order $\mathcal{O}(\partial)$ in Eqs.~(\ref{eq:alphadot})-(\ref{eq:udot}). However, for completeness, we nevertheless discuss here the result for the first-order
terms in gradient expansion. The results of the present section are the same as in Ref.~\cite{Jaiswal:2013npa}
which was derived for zero magnetic fields.

We evaluate the dissipative part of the energy-momentum tensor (which includes the shear,
bulk viscosity, and diffusion) using $\delta \tilde{f}^{(1)}$ and $\delta \tilde{\bar{f}}^{(1)}$ in the following,
\begin{eqnarray}
\label{firstshear}
  \pi^{\mu\nu}_{(1)}&=&\Delta^{\mu\nu}_{\alpha\beta}\int dp p^{\alpha}p^{\beta}\left(\delta \tilde{f}^{(1)}+\delta \tilde{\bar{f}}^{(1)}\right),\\
\label{firstbulk}
  \Pi_{(1)}&=&-\frac{\Delta_{\mu\nu}}{3}\int dp p^{\mu}p^{\nu}\left(\delta \tilde{f}^{(1)}+\delta \tilde{\bar{f}}^{(1)}\right),\\
\label{firstdiffusion}
  V^{\mu}_{(1)}&=&\Delta^{\mu}_{\alpha}\int dp p^{\alpha}\left(\delta \tilde{f}^{(1)}-\delta \tilde{\bar{f}}^{(1)}\right).
\end{eqnarray}
Substituting the value of $\delta \tilde{f}^{(1)}$ from Eq.~\eqref{eq:deltaf1} into the above Eqs.~\eqref{firstshear}-\eqref{firstdiffusion}, after some algebra we get the following relations. For shear viscous pressure
\begin{equation}\label{resultshear1}
\pi^{\mu\nu}_{(1)}=2 \tau_c \beta_{\pi}\sigma^{\mu\nu},
\end{equation}
where $\beta_{\pi}=\beta J_{42}^{(1)+}$ and $\sigma^{\mu\nu}=\Delta^{\mu\nu}_{\alpha\beta}\nabla^{\alpha}u^{\beta}$.

For the bulk viscous pressure,
\begin{equation}\label{resultbulk1}
  \Pi_{(1)}=-\tau_c \beta_{\Pi}\theta,
\end{equation}
where $\theta=\partial_{\mu}u^{\mu}$ and
\begin{equation}
\beta_{\Pi}=\frac{5\beta}{3}J_{42}^{(1)+}+\mathcal{X} J_{31}^{(0)+}-\mathcal{Y} J_{21}^{(0)-},
\end{equation}
with the terms $\mathcal{X}$ and $\mathcal{Y}$ being
\begin{eqnarray}
{\label{lm}}
 \nonumber
  \mathcal{X} &=& \frac{J_{10}^{(0)+}(\epsilon+P)-J_{20}^{(0)-}n_{f}}{D_{20}}, \\
  \mathcal{Y} &=& \frac{J_{20}^{(0)-}(\epsilon+P)-J_{30}^{(0)+}n_{f}}{D_{20}},
\end{eqnarray}
respectively.

Finally for the net particle diffusion current,
\begin{equation}\label{resultdiffusion1}
  V^{\mu}_{(1)}=\tau_c \beta_V \nabla^{\mu}\alpha,
\end{equation}
where $\beta_V=\frac{n_{f}}{\epsilon+P} J_{21}^{(0)-}-J_{21}^{(1)-}$.
\subsection{Second order equation}
We derive the second-order relaxation type equations for the shear, bulk viscous pressure
and diffusion current by taking the appropriate moments of $\delta \tilde{f}^{(2)}$.
While deriving these equations, we keep terms up to order $\mathcal{O}(\partial^2)$.
We know that the second-order transport coefficient differs even for zero magnetic fields
when calculated using RTA~\cite{Jaiswal:2013npa} and moment method~\cite{Betz:2009zz}. We might expect a similar result for non-zero
magnetic field as well.
\subsubsection{Evolution for the shear stress}
By definition the second order contribution to the shear stress tensor is given by:


\begin{equation}
\label{eq:shear2}
  \pi^{\mu\nu}_{(2)} = \Delta^{\mu\nu}_{\alpha\beta}\int dp p^{\alpha}p^{\beta} \left(\delta \tilde{f}^{(2)}+\delta \tilde{\bar{f}}^{(2)}\right),
\end{equation}

where $\delta \tilde{f}^{(2)}$ is given in Eq.~\eqref{eq:deltaf2}. Note that the total shear stress is
the combination of first and second order terms:
\begin{equation}\label{2evolutionstress}
  \pi^{\mu\nu}=\pi^{\mu\nu}_{(1)} +\pi^{\mu\nu}_{(2)}.
\end{equation}
Evaluating the integral of Eq.~\eqref{eq:shear2} (see Appendix~(\ref{app:shear}) for details) and adding
it to the Eq.~\eqref{2evolutionstress} we get the evolution equation for the shear stress tensor:

\begin{widetext}
\begin{eqnarray}
\label{eq:shear_evolution2}
\nonumber
  \frac{\pi^{\mu\nu}}{\tau_c}&=&-\dot{\pi}^{\mu\nu}+2\beta_{\pi}\sigma^{\mu\nu}+2\pi^{\langle\mu}_{\gamma}\omega^{\nu\rangle\gamma}-\tau_{\pi\pi}\pi^{\langle\mu}_{\gamma}\sigma^{\nu\rangle\gamma} -\delta_{\pi\pi}\pi^{\mu\nu}\theta +\lambda_{\pi\Pi}\Pi\sigma^{\mu\nu}-\tau_{\pi V}V^{\langle\mu}\dot{u}^{\nu\rangle}
   +\lambda_{\pi V}V^{\langle\mu}\nabla^{\nu \rangle}\alpha +l_{\pi V}\nabla^{\langle\mu}V^{\nu\rangle} \\
   &+&\delta_{\pi B}\Delta^{\mu\nu}_{\eta \beta}q B b^{\gamma \eta}g^{\beta \rho}\pi_{\gamma\rho}
  -\tau_c qB\tau_{\pi VB} \dot{u}^{\langle\mu}b^{\nu\rangle\sigma} V_{\sigma}-\tau_c qB\lambda_{\pi VB} V_{\gamma}b^{\gamma\langle\mu}\nabla^{\nu\rangle}\alpha-q\tau_c \delta_{\pi VB}   \nabla^{\langle\mu}\left(B^{\nu\rangle\gamma}V_{\gamma} \right),
\end{eqnarray}
\end{widetext}
the resulting second order transport co-efficients are given in terms of thermodynamic integrals in Table.~\eqref{table:ShearJmn}. Note, that the co-efficients $\tau_{\pi V}$ and $\lambda_{\pi V}$ contain the derivatives of $l_{\pi V}$ \cite{Denicol:2012es}, while $\tau_{\pi VB},\lambda_{\pi VB}$ contain derivatives of $\delta_{\pi V B}$ respectively.
 We notice that the last four terms contain the magnetic field explicitly and
 are new  when compared to the case for zero magnetic field Ref.~\cite{Jaiswal:2013npa}. 
 Compared to the calculation done for non-zero magnetic field using a 14-moment 
 approximation in Ref.~\cite{Denicol:2018rbw}, we found only the first
 ten terms have a similar form albeit, different coefficients. However, the last four
 terms are new and do not appear in the 14- moment approximation. We will discuss
 this issue in Sec.~\ref{sec:discuss}.
\subsubsection{Evolution for the bulk stress}
Similar to the shear viscosity, we derive the second order evolution equation for
the bulk viscous stress. By the definition:
\begin{eqnarray}
\label{eq:bulk2}
 \Pi_{(2)}=-\frac{\Delta_{\alpha\beta}}{3}\int dp p^{\alpha}p^{\beta}\left(\delta \tilde{f}^{(2)}+\delta \tilde{\bar{f}}^{(2)}\right).
 \end{eqnarray}
Evaluating the above integral by using $\delta \tilde{f}^{(2)}$ from Eq.~\eqref{eq:deltaf2} and
noting the fact that the total bulk stress is a combination of first and second-order terms i.e.,
$\Pi=\Pi_{(1)}+\Pi_{(2)}$ after some algebra (the details are given in Appendix~\ref{AppSec:bulk} we get the
evolution equation for bulk stress:
\begin{widetext}
\begin{eqnarray}
\nonumber
   \frac{\Pi}{\tau_c}&=&-\dot{\Pi}-\delta_{\Pi\Pi}\Pi \theta +\lambda_{\Pi\pi}\pi^{\mu\nu}\sigma_{\mu\nu}-\tau_{\Pi V}V\cdot \dot{u}-\lambda_{\Pi V}V\cdot \nabla \alpha
    -l_{\Pi V}\partial \cdot V-\beta_{\Pi}\theta+\tau_c \tau_{\Pi V B}\dot{u}_{\alpha} qBb^{\alpha\beta}V_{\beta}\\
    &-&\tau_c q \delta_{\Pi V B}\nabla_{\mu}\left(B b^{\mu\beta}V_{\beta} \right)-\tau_c qB\lambda_{\Pi VB} b^{\mu\beta}V_{\beta}\nabla_{\mu}\alpha,
    \label{eq: bulk_evolution2}
\end{eqnarray}
\end{widetext}
where the second-order transport coefficients are given in terms of the thermodynamic integrals in
Table.~\ref{table:BulkJmn} and we use Eq.~\eqref{lm} for the expression of $\mathcal{X}$ and $\mathcal{Y}$. Coefficients $\tau_{\Pi V},\lambda_{\Pi V}$ contain derivatives of $l_{\Pi V}$, while $\tau_{\Pi VB},\lambda_{\Pi VB}$ contain derivatives of $\delta_{\Pi V B}$, respectively. The last three terms of the above equation are new compared to that of \cite{Jaiswal:2013npa} and are magnetic field dependent. When compared to
the 14-moment approximation \cite{Denicol:2018rbw} in the presence of a magnetic field,
the bulk viscous relaxation equation did not have any magnetic field dependent term.
\begin{table*}
\label{table:1}
\begin{center}
\begin{tabular}{|p{2cm}|p{1.3cm}|p{1.3cm}|p{1.3cm}|p{1.3cm}|p{1.3cm}|p{1.3cm}|}
 \hline
 \text{}&$\beta_{\pi}$&$\tau_{\pi\pi}$& $\delta_{\pi\pi}$ & $\tau_{\pi V}$& $\lambda_{\pi V}$& $l_{\pi V}$\\
 \hline
  \text{Denicol et al. }&$4P/5$&$10/7$& $4/3$ & $0$& $0$& $0$\\
 \hline
 \text{CE}&$4P/5$&$10/7$& $4/3$ & 0& $0$& $0$ \\
  \hline
 \end{tabular}
\\[10pt]
\caption{Comparison between the coefficients for the shear-stress equation for a massless Boltzmann gas (here we compare
the result for particles only) calculated in this work using CE method and Denicol et al. using the 14-moment method~\cite{Denicol:2019iyh}.}
\label{table:ShearSecondOrder}
\end{center}
\end{table*}
\begin{table*}
\begin{center}
\begin{tabular}{|p{2cm}|p{1.3cm}|p{1.3cm}|p{1.3cm}|p{1.3cm}|p{1.3cm}|p{1.3cm}|}
 \hline
 \text{}&$\beta_{V}$&$\lambda_{VV}$& $\delta_{VV}$ & $\tau_{V \pi }$& $\lambda_{V \pi }$& $l_{V \pi}$ \\
 \hline
  \text{Denicol et al.}&$n_f/12$&$3/5$& $1$ & $\beta/20$& $\beta/20$& $\beta/20$\\
 \hline
   \text{CE}&$n_f/12$&$3/5$& $1$ & $\beta/4$& $\beta/4$& $\beta/4$\\
 \hline
 \end{tabular}
\\[10pt]
\caption{Comparison between the coefficients for the diffusion equation for a massless Boltzmann gas calculated in this work using Chapman-Enskog method (CE) and Denicol et al. using the 14-moment method~\cite{Denicol:2019iyh} (particle only).}
\label{table:DiffusionSecondOrder}
\end{center}
\end{table*}


\begin{table*}
\begin{center}
\begin{tabular}{|p{2cm}|p{1.22cm}|p{1.22cm}|p{1.22cm}|p{1.22cm}|p{1.22cm}|p{1.22cm}|p{1.22cm}|p{1.22cm}|p{1.22cm}|p{1.22cm}|p{1.22cm}|p{1.22cm}|}
 \hline
 \text{}&$\delta_{\pi B}$&$\delta_{VB}$&$\delta_{\pi VB}$&$\delta_{\Pi VB}$& $\tau_{\Pi VB}$ & $l_{V\pi B}$&$\tau_{V\Pi B}$& $l_{V\Pi B}$&$\delta_{VVB}$&$\lambda_{VV B}$&$\rho_{VVB}$&$\tau_{VVB}$\\
 \hline
\text{Denicol et al.}&$2\beta/5$&$5\beta/12$& $-$ & $-$& $-$& $-$&$-$&$-$&$-$&$-$&$-$&$-$\\
   \hline
 \text{CE}&$\beta/2$&$\beta$&$2/5$&$1/3$& $2/3$ & $\beta^2/12$& $\beta^2/12$& $\beta^2/12$&$\beta/3$&$3\beta/20$&$\beta/4$&$\beta/4$ \\
  \hline
 \end{tabular}
\\[10pt]
\caption{Transport coefficients appearing in the shear, bulk and diffusion equation that couple magnetic field and dissipative quantities for a massless Boltzmann gas (particles only).}
\label{table:ThirdOrder}
\end{center}
\end{table*}

\subsubsection{Diffusion current}
The expression for the diffusion current for the net charge in second order is:
\begin{equation}
\label{eq:Diffusion2}
  V^{\mu}_{(2)}= \Delta^{\mu}_{\alpha}\int dp p^{\alpha} \left(\delta \tilde{f}^{(2)}-\delta \tilde{\bar{f}}^{(2)} \right),
\end{equation}
where $\delta\tilde{f}_{(2)}$ is taken from Eq.~\eqref{eq:deltaf2}.
Like other dissipative quantities, the total diffusion four vector is composed of
first and second order terms, i.e., $V^{\mu}=V^{\mu}_{(1)}+V^{\mu}_{(2)}$.
After evaluating the integral (for details see Appendix~\ref{AppSubsec:Diffusion} in Eq.~\eqref{eq:Diffusion2},
we get the following second order evolution equation for the diffusion current:
\begin{widetext}
\begin{eqnarray}
\label{eq:diffusionEvolution2}
\nonumber
  \frac{V^{\mu}}{\tau_c}&=&-\dot{V}^{\langle\mu\rangle}-V_{\nu}\omega^{\nu \mu}-\lambda_{VV}V^{\nu}\sigma^{\mu}_{\nu}-\delta_{VV}V^{\mu} \theta+\lambda_{V\Pi}\Pi\nabla^{\mu} \alpha
  -\lambda_{V\pi}\pi^{\mu \nu}\nabla_{\nu}\alpha -\tau_{V\pi}\pi^{\mu}_{\nu}\dot{u^{\nu}}+\tau_{V\Pi}\Pi \dot{u^{\mu}}+l_{V\pi}\Delta^{\mu \nu}\partial_{\gamma}\pi^{\gamma}_{\nu} \\
  \nonumber
  &&-l_{V\Pi}\nabla^{\mu}\Pi+\beta_V \nabla^{\mu} \alpha-q B \delta_{V B} b^{\mu\gamma}V_{\gamma} +\tau_c q B l_{V\pi B} b^{\sigma \mu}\partial^{\kappa}\pi_{\kappa\sigma}+\tau_c q B \tau_{V \Pi B}b^{\gamma\mu}\Pi \dot{u}_{\gamma} -\tau_c qB l_{V \Pi B} b^{\gamma\mu}\nabla_{\gamma}\Pi\\
  &-&q\tau_c \delta_{VVB} B b^{\mu\nu}V_{\nu}\theta-q \tau_c \lambda_{VVB}B b^{\gamma \nu}V_{\nu}\sigma^{\mu}_{\gamma}-q \tau_c \mathbf{\rho}_{VVB} B b^{\gamma \nu}V_{\nu} \omega^{\mu}_{\gamma}  -\tau_c q \tau_{VVB}\Delta^{\mu}_{\gamma}D\left(Bb^{\gamma \nu}V_{\nu} \right)
\end{eqnarray}
\end{widetext}
where the second order transport coefficients are given in terms of thermodynamic integrals in Table.~\ref{table:DiffusionJmn}. Coefficients $\tau_{V\pi}$, $\lambda_{V\pi}$  contain the derivative of $l_{V\pi}$; $\tau_{V\Pi}$, $\lambda_{V\Pi}$  contain the derivative of $l_{V\Pi}$ and $\delta_{VVB}$ contains derivative of $\tau_{VVB}$,  respectively.
To arrive at the final expression Eq.~\eqref{eq:diffusionEvolution2} we also make use of
$\mathcal{X}$ and $\mathcal{Y}$ given in Eq.~\eqref{lm}. A comparison of the above to
the RTA calculation done in \cite{Jaiswal:2013npa} without magnetic field shows
 that the last eight terms are new and are magnetic field dependent. A similar
 comparison with the relaxation equation for the diffusion in the presence of a
 magnetic field derived in \cite{Denicol:2018rbw} using 14-moment approximation
 shows that only the first twelve terms are of similar form, while the last seven
 terms are not present in the moment method.
\subsection{Ultrarelativistic and the weak field limit}
{\label{sec:discuss}}
The transport coefficients in the ultra-relativistic limit, i.e. $m/T=0$, for a classical
Maxwell gas with a constant relaxation time $\tau_c$, can be calculated analytically
using the thermodynamic integrals. The transport coefficients are grouped into:
(i) those which are independent of the magnetic field are collected in
Tables.~\ref{table:ShearSecondOrder} and \ref{table:DiffusionSecondOrder}
for the shear and the number diffusion respectively. (ii) Those which are magnetic field
dependent are collected in Table.~\ref{table:ThirdOrder}. In this limit, the bulk
viscous pressure vanishes and has not been considered. In the same table, the
results from the 14-moment approximation in the presence of a magnetic field
\cite{Denicol:2018rbw} in the ultra-relativistic limit have also been shown.
It is worthwhile to note that in this limit, the new coefficients namely
$\delta_{\pi B}$ and $\delta_{VB}$ are different in the above two approaches.

In the limit of weak magnetic field, which translates to the statement that temperature
of  the  system  is sufficiently large than the strength of the magnetic field $T^2\gg qB$.
We define the dimensionless parameter $g_B=qB/T^2$ such that $g_B\ll 1$.
The RTA approximation in the presence of  magnetic field Eq.~\eqref{eq:expansion},
has two power counting schemes, viz. $\mathrm{Kn}=\tau_c T$ and $\chi=qB\tau_c/T$.
However, in the weak field limit, the expansion parameter $\chi=g_B\tau_c T$ becomes
smaller and hence treated as sub-leading contribution. Therefore, at second order one effectively retains
term till $\mathcal{O}(\mathrm{Kn}^2)$ in spatial gradients and
$\mathcal{O}(\chi\cdot \mathrm{Kn})$ for the mixed terms~\footnote{We do not keep terms which are $\mathcal{O}(\chi^2)$, since they do not contribute to the expansion Eq.~\eqref{eq:expansion}. }.
In this limit the relaxation equations reduce to following forms:
\begin{widetext}
\begin{eqnarray}
\label{eq:shear_evolution3}
\nonumber
\dot{\pi}^{\mu\nu}  &=&2\beta_{\pi}\sigma^{\mu\nu}-\frac{\pi^{\mu\nu}}{\tau_c}+2\pi^{\langle\mu}_{\gamma}\omega^{\nu\rangle\gamma}-\tau_{\pi\pi}\pi^{\langle\mu}_{\gamma}\sigma^{\nu\rangle\gamma} -\delta_{\pi\pi}\pi^{\mu\nu}\theta +\lambda_{\pi\Pi}\Pi\sigma^{\mu\nu}-\tau_{\pi V}V^{\langle\mu}\dot{u}^{\nu\rangle}
   +\lambda_{\pi V}V^{\langle\mu}\nabla^{\nu \rangle}\alpha   \\
   &&+l_{\pi V}\nabla^{\langle\mu}V^{\nu\rangle}+\delta_{\pi B}\Delta^{\mu\nu}_{\eta \beta}q B b^{\gamma \eta}g^{\beta \rho}\pi_{\gamma \rho}, \\
       \label{eq: bulk_evolution3}
 \dot{\Pi}  &=&-\beta_{\Pi}\theta-\frac{\Pi}{\tau_c}-\delta_{\Pi\Pi}\Pi \theta +\lambda_{\Pi\pi}\pi^{\mu\nu}\sigma_{\mu\nu}-\tau_{\Pi V}V \cdot \dot{u}-\lambda_{\Pi V}V\cdot \nabla \alpha
    -l_{\Pi V}\partial \cdot V,\\
    \label{eq:diffusionEvolution3}
\nonumber
  \dot{V}^{\langle\mu\rangle}&=&\beta_V \nabla^{\mu} \alpha-\frac{V^{\mu}}{\tau_c}-V_{\nu}\omega^{\nu \mu}-\lambda_{VV}V^{\nu}\sigma^{\mu}_{\nu}-\delta_{VV}V^{\mu} \theta+\lambda_{V\Pi}\Pi\nabla^{\mu} \alpha
  -\lambda_{V\pi}\pi^{\mu \nu}\nabla_{\nu}\alpha -\tau_{V\pi}\pi^{\mu}_{\nu}\dot{u^{\nu}}+\tau_{V\Pi}\Pi \dot{u^{\mu}}\\
  &&+l_{V\pi}\Delta^{\mu \nu}\partial_{\gamma}\pi^{\gamma}_{\nu}-l_{V\Pi}\nabla^{\mu}\Pi-q B \delta_{V B} b^{\mu\gamma}V_{\gamma}.
\end{eqnarray}
\end{widetext}
\begin{figure}
\centering
\includegraphics[height =0.34\textwidth]{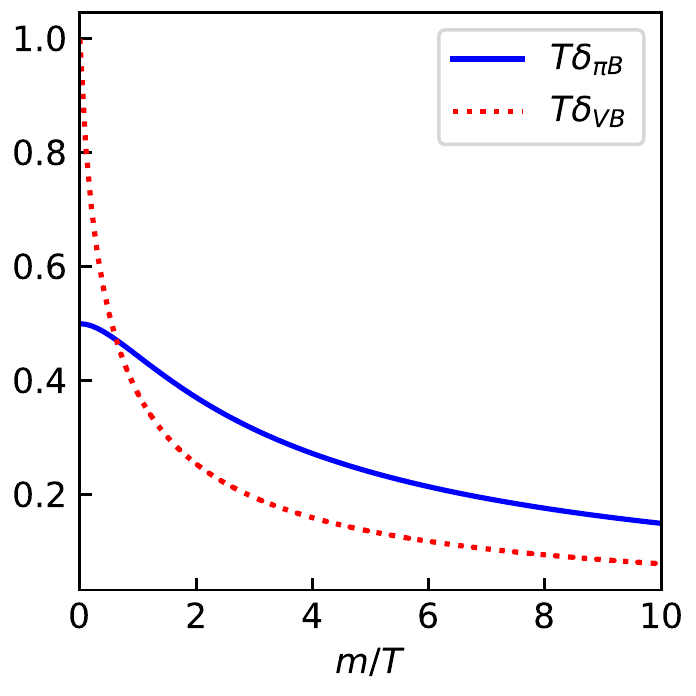}
\caption{Dimensionless transport coefficients $T \delta_{VB}$ and $T\delta_{\pi B}$ that couple fluid to magnetic field as a function of ${m}/{T}$.}
\label{fig:TransMag}
\end{figure}
The above set of relaxation equations, now have the same form as that of Ref.~\cite{Denicol:2018rbw}. The dimensionless magnetic field dependent transport coefficients $T\delta_{\pi B}$ and $T\delta_{VB}$ in the weak field limit are plotted in Fig.~\ref{fig:TransMag} as a function of $m/T$. In the limit $m/T\rightarrow 0$, these coefficients reduce to those obtained in Tables.~\ref{table:ThirdOrder}.

\subsection{The Navier-Stokes limit}
In the Navier-Stokes limit, we keep terms  $\mathcal{O}(\mathrm{Kn})$,
and $\mathcal{O}(\chi\cdot \mathrm{Kn})$, which leaves us with the first and second
terms in right hand side of Eqs.~\eqref{eq:shear_evolution3}-\eqref{eq:diffusionEvolution3}
which are of first-order in gradients as well as the last term which is magnetic field
dependent and are also first-order in gradients. Bringing these terms to the left we have,
\begin{eqnarray}
\label{eq: bulk_evolution3}
   \frac{\Pi}{\tau_c}&=&\beta_{\Pi}\theta,\\
   \label{eq:diffusionEvolution4}
  \left(\frac{g^{\mu\nu}}{\tau_c}+q B \delta_{V B} b^{\mu\gamma}\right)V_{\gamma}&=&\beta_V \nabla^{\mu} \alpha,\\
\label{eq:shear_evolution4}
  \left(\frac{g^{\mu\gamma}g^{\nu\rho}}{\tau_c}-\delta_{\pi B}\Delta^{\mu\nu}_{\eta \beta}q B b^{\gamma \eta}g^{\beta \rho}\right)\pi_{\gamma\rho}&=&2\beta_{\pi}\sigma^{\mu\nu}.
\end{eqnarray}

Since the bulk viscous pressure does not involve any magnetic field dependence,
the Navier-Stokes limit for bulk viscosity $\zeta$ turns out to be trivially the same
as that without any magnetic field, i.e., $\zeta=\beta_{\Pi}\tau_c$. One needs now
to invert the coefficients multiplied in the left of the rest of the equations to get the
respective constitutive relations. The general solution for the rest of the equations
are given as
\begin{widetext}
\begin{eqnarray}
 V_{\gamma}&=&\left(\kappa_{\parallel}P^{\parallel}_{\delta\gamma}+\kappa_{\bot}P^{\bot}_{\delta\gamma}+\kappa_{\times}P^{\times}_{\delta\gamma}\right)\partial^{\delta}\alpha,
 \\
 \pi_{\gamma\rho}&=&\bigg(\eta_0P^{(0)}_{\alpha\beta\gamma\rho}+\eta_1\left(P^{(1)}_{\alpha\beta\gamma\rho}+P^{(-1)}_{\alpha\beta\gamma\rho}\right)+i\eta_2\left(P^{(1)}_{\alpha\beta\gamma\rho}-P^{(-1)}_{\alpha\beta\gamma\rho}\right)
 +\eta_3\left(P^{(2)}_{\alpha\beta\gamma\rho}+P^{(-2)}_{\alpha\beta\gamma\rho}\right)+i\eta_4\left(P^{(2)}_{\alpha\beta\gamma\rho}-P^{(-2)}_{\alpha\beta\gamma\rho}\right)\bigg)\sigma^{\alpha\beta}.\nonumber\\
\end{eqnarray}
\end{widetext}
where $P^{\parallel}_{\delta\gamma}$, $P^{\bot}_{\delta\gamma}$ and $P^{\times}_{\delta\gamma}$ are second rank projection tensors while $P^{(n)}_{\alpha\beta\gamma\rho}$, with $n=-2$ to $n=+2$ are fourth rank projection tensors respectively. The definition of the these tensors can be found in Ref.~\cite{Hass}. In the above set of equations $\kappa_{\parallel}$, $\kappa_{\bot}$, $\kappa_{\times}$ and $\eta_{0}-\eta_{4}$ are the transport coefficients. These coefficients can be obtained by substituting the above solution to the left hand side of Eq.~\eqref{eq:shear_evolution4} and Eq.~\eqref{eq:diffusionEvolution4} and using the usual properties of projection tensors. The resulting transport coefficients can be written as
\begin{eqnarray}
 \kappa_{\parallel}&=&\beta_{V}\tau_c,\\
\kappa_{\bot}&=&\frac{\beta_{V}\tau_c}{1+{\left(qB\tau_c\delta_{VB}\right)}^2},\\
\kappa_{\times}&=&\frac{\beta_{V} qB\tau_c^2\delta_{VB}}{1+{\left(qB\tau_c\delta_{VB}\right)}^2}=\kappa_{\bot}qB\tau_c\delta_{VB},
\end{eqnarray}
for the diffusion coefficients and
\begin{eqnarray}
 \eta_0&=&2\beta_{\pi}\tau_c,\\
  \eta_1&=&\frac{8\beta_{\pi}\tau_c}{4+{\left(qB\tau_c\delta_{\pi B}\right)}^2},\\
  \eta_2&=&\frac{4\beta_{\pi}qB\tau_c^2\delta_{\pi B}}{4+{\left(qB\tau_c\delta_{\pi B}\right)}^2}=\frac{\eta_1 qB\tau_c\delta_{\pi B}}{2},\\
  \eta_3&=&\frac{2\beta_{\pi}\tau_c}{1+{\left(qB\tau_c\delta_{\pi B}\right)}^2},\\
  \eta_4&=&\frac{2\beta_{\pi}qB\tau_c^2\delta_{\pi B}}{1+{\left(qB\tau_c\delta_{\pi B}\right)}^2}={\eta_3 qB\tau_c\delta_{\pi B}},
\end{eqnarray}
for the shear viscousty, respectively.

In the limit of vanishing magnetic field, i.e., $qB\rightarrow 0$, the diffusion
coefficients reduce to $\kappa_{\times}\rightarrow 0$ and $\kappa_{\parallel}=\kappa_{\bot}$,
while the shear viscous coefficients reduce to $\eta_2=\eta_4=0$ and $\eta_1=\eta_3=\eta_0$
as expected.

\section{Conclusion}
{\label{sec:conclusion}}
We derive for the first time the relativistic non-resistive, viscous second-order magnetohydrodynamics equations for the dissipative quantities using the relaxation time approximation. Assuming that the single-particle distribution function is close to equilibrium, we solve the Boltzmann equation in the presence of a magnetic field
using Chapman-Enskog like gradient expansion with two relevant expansion parameters: the Knudsen number and a dimensionless parameter $\chi=qB\tau_c/T$ that depends on the strength of the magnetic field.
In first-order, dissipative quantities are found to be independent of the magnetic field. Moreover, in second-order, we found new transport coefficients that couple magnetic field to dissipative quantities apart from the usual transport coefficients that one gets without any external field.
When compared to the results of the 14-moment approximation, additional terms 
involving the magnetic field appear in the relaxation time approximation. However, 
in the weak field limit, the form of the relaxation equations is the same as that of 
the 14-moment approximation but with different values for the transport coefficients. 
In the ultra-relativistic limit, the resulting transport coefficients from the two approaches 
are compared, some of the coefficients are found to differ.
Finally, we find that one recovers the usual anisotropic transport coefficients for fluid in
magnetic fields in the Navier-Stokes limit. As a further extension of the present work, 
we plan to investigate the general case for the relativistic resistive viscous fluid in
a magnetic field in a future study~\cite{rta:res}.The
formulation of the relativistic causal magnetohydrodynamics is expected to be useful
in astrophysical phenomena involving relativistic plasmas as well as femto-scopic
high energy heavy-ion collisions.

\begin{acknowledgments}
AP acknowledges the CSIR-HRDG financial support. AD, and VR acknowledges support from the DAE, Govt. of India. RB and VR acknowledge
financial support from the DST Inspire faculty research grant (IFA-16-PH-167), India.
\end{acknowledgments}
\onecolumngrid
\appendix

\section{Thermodynamic Integrals}{\label{appendix1}}
The $n$-th moments integral for the distribution function is defined as:
\begin{eqnarray}
 I_{\mu_{1}\mu_{2}\cdots\mu_{n}}^{(m)\pm}=\int \frac{dp}{\left(u\cdot p\right)^{m }}p_{\mu_{1}}p_{\mu_{2}}\cdots p_{\mu_{n}} \left(f_{0}\pm \bar{f_0}\right),
\end{eqnarray}
which can be docomposed as:
\begin{equation}
\label{eq:Iexpansion}
I_{\mu_{1}\mu_{2}\dots\mu_{n}}^{(m)\pm}=I_{n0}^{(m)\pm}u_{\mu_{1}}\cdots u_{\mu_{n}}+I_{n1}^{(m)\pm}\left(\Delta_{\mu_{1}\mu_{2}} u_{\mu_{3}} \cdots u_{\mu_{n}}+ \text{perm.}\right)+ \dots + I_{nq}^{(m)\pm}\left(\Delta_{\mu_{1}\mu_{2}}\Delta_{\mu_{3}\mu_{4}}\cdots \Delta_{\mu_{n-1}\mu_{n}} +\text{perm.} \right).
\end{equation}
where $n\geq 2q$.\par
Similarly the auxiliary moments integral
\begin{eqnarray}
 J_{\mu_{1}\mu_{2}\cdots \mu_{n}}^{(m)\pm}=\int \frac{dp}{\left(u\cdot p\right)^{m}}p_{\mu_{1}}p_{\mu_{2}}\cdots p_{\mu_{n}} \left(f_{0}\tilde{f_{0}}\pm \bar{f_{0}}\tilde{\bar{f}}_{0} \right),
\end{eqnarray}
can be decomposed as:
\begin{equation}
\label{Jexpansion}
J_{\mu_{1}\mu_{2}\dots\mu_{n}}^{(m)\pm}=J_{n0}^{(m)\pm}u_{\mu_{1}}\cdots u_{\mu_{n}}+J_{n1}^{(m)\pm}\left(\Delta_{\mu_{1}\mu_{2}} u_{\mu_{3}} \cdots u_{\mu_{n}}+ \text{perm.}\right)+ \dots + J_{nq}^{(m)\pm}\left(\Delta_{\mu_{1}\mu_{2}}\Delta_{\mu_{3}\mu_{4}}\cdots \Delta_{\mu_{n-1}\mu_{n}} +\text{perm.} \right).
\end{equation}
where $\tilde{f}_0=1-rf_0$.
Here we define the thermodynamic integrals as follows:
\begin{equation}
\label{eq:I}
  I^{(m)\pm}_{nq}=\frac{1}{(2q+1)!!}\int_{ }^{}dp (u \cdot p)^{n-2q-m}(\Delta_{\alpha\beta}p^{\alpha}p^{\beta})^q \left(f_0\pm \bar{f_0}\right),
\end{equation}
and
\begin{equation}
\label{eq:J}
  J^{(m)\pm}_{nq}=\frac{1}{(2q+1)!!}\int_{ }^{}dp (u \cdot p)^{n-2q-m}(\Delta_{\alpha\beta}p^{\alpha}p^{\beta})^q \left(f_0 \tilde{f}_0 \pm \bar{f}_0 \tilde{\bar{f}}_0\right).
\end{equation}
One can write the $J$ in terms of $I$ as:
\begin{equation}\label{relJI}
  J^{(0)\pm}_{nq}=\frac{1}{\beta}\left[-I^{(0)\pm}_{n-1,q-1}+(n-2q)I^{(0)\pm}_{n-1,q}\right].
\end{equation}

The general expression of $D_{nq}$ used in Eq.~\eqref{eq:alphadot} and Eq.~\eqref{eq:betadot}
is given by: $D_{nq}=J_{n+1,q}^{(0)+}J_{n-1,q}^{(0)+}-J_{nq}^{(0)-}J_{nq}^{(0)-}$.

\section{Second order relaxation equation for dissipative stresses}{\label{appendix2}}
In this appendix we discuss the detail calculation of the second order dissipative stresses.
The contribution due to the antiparticles are not shown explicitly for simplicity
but they appear in the final expressions.

\subsection{Shear stress}
\label{app:shear}
The second order shear stress $\pi^{\mu\nu}_{(2)}$ is given by Eq.~\eqref{eq:shear2}:
\begin{eqnarray}
\pi^{\mu\nu}_{(2)}=\Delta^{\mu\nu}_{\alpha\beta}\int_{}^{}dp p^{\alpha}p^{\beta}\left(\frac{\tau_c}{u \cdot p}p^{\rho} \partial_{\rho} \left[\frac{\tau_c}{u \cdot p}p^{\sigma} \partial_{\sigma} f_0\right]+\frac{\tau_c}{u \cdot p}qF^{\gamma\eta}p_{\eta}\frac{\partial}{\partial p^{\gamma}}\left[\frac{\tau_c}{u \cdot p}p^{\sigma}\partial_{\sigma}f_0\right]\right).
\end{eqnarray}
For convenience, we write them into two parts as:
\begin{equation}\label{IandII}
  \pi^{\mu\nu}_{(2)} = \mathcal{I}_{1}+\mathcal{I}_{2}.
\end{equation}
Here
\begin{eqnarray}\label{eq:Iexpression}
  \mathcal{I}_{1}&=& \Delta^{\mu\nu}_{\alpha\beta}\int_{}^{}dp p^{\alpha}p^{\beta}\left(\frac{\tau_c}{u \cdot p}p^{\rho} \partial_{\rho} \left[\frac{\tau_c}{u \cdot p}p^{\sigma} \partial_{\sigma} f_0\right]\right),\\
   \label{IIshear}
   \mathcal{I}_{2}&=& \Delta^{\mu\nu}_{\alpha\beta}\int_{}^{}dp p^{\alpha}p^{\beta} \left(\frac{\tau_c}{u \cdot p}qF^{\gamma\eta}p_{\eta}\frac{\partial}{\partial p^{\gamma}}\left[\frac{\tau_c}{u \cdot p}
  p^{\sigma}\partial_{\sigma}f_0\right]\right),
\end{eqnarray}
Let us first evaluate the integral $\mathcal{I}_{1}$:
\begin{eqnarray}
\nonumber
\mathcal{I}_{1}&=&\Delta^{\mu\nu}_{\alpha \beta}\int_{}^{}dp p^{\alpha}p^{\beta} \left(\frac{\tau_c}{u \cdot p}p^{\rho} \partial_{\rho} \left[\frac{\tau_c}{u \cdot p}p^{\sigma} \partial_{\sigma} f_0\right]\right),\\
                       &=& \mathcal{A}+\mathcal{B}+\mathcal{C},
\label{abc}
\end{eqnarray}
where
\begin{eqnarray}
\nonumber
\mathcal{A}&=&\Delta^{\mu\nu}_{\alpha \beta}\int_{}^{}dp p^{\alpha}p^{\beta} \tau_c D\left[\frac{\tau_c}{u \cdot p}p^{\sigma}\partial_{\sigma} f_0\right],\\
\nonumber
\mathcal{B}&=&\Delta^{\mu\nu}_{\alpha \beta}\int_{}^{}dp p^{\alpha}p^{\beta} \frac{\tau_c}{u \cdot p}p^{\rho}\nabla_{\rho}\left[\tau_c \dot{f_0}\right],\\
\nonumber
\mathcal{C}&=&\Delta^{\mu\nu}_{\alpha \beta}\int_{}^{}dp p^{\alpha}p^{\beta} \frac{\tau_c}{u \cdot p}p^{\rho}\nabla_{\rho} \left[\frac{\tau_c}{u \cdot p}p^{\sigma}\nabla_{\sigma} f_0\right].
\end{eqnarray}
A straight forward calculation gives:
\begin{eqnarray}
\nonumber
\mathcal{A}&=&-\Delta^{\mu\nu}_{\alpha \beta}\int_{}^{}dpf_0 \tilde{f}_0 p^{\alpha}p^{\beta} \tau_c D\left[\frac{\tau_c}{u \cdot p}p^{\sigma}\left\{\ \beta p^{\gamma}\partial_{\sigma}u_{\gamma}+\left(u \cdot p\right)\partial_{\sigma}\beta-\partial_{\sigma}\alpha\right\}\ \right].
\end{eqnarray}
We can rewrite the above expression in terms of the thermodynamic integrals given in Appendix~\ref{appendix1} and
Eq.~\eqref{eq:udot} as:
\begin{eqnarray}
\label{eq:a}
  \mathcal{A}&=&-\tau_c \dot{\pi}^{\langle\mu\nu\rangle}-2\tau_c^2 \left(\frac{n_f}{\epsilon+P}J_{31}^{(0)-}-J_{31}^{(1)-}\right)\dot{u}^{\langle\mu}\nabla^{\nu\rangle}\alpha
  +\tau_c^2\Delta^{\mu\nu}_{\alpha \beta}J_{31}^{(0)-}\dot{u}^{\beta}\left[\frac{\beta qBb^{\alpha \sigma}}{\epsilon+P}V_{\sigma}\right]+\tau_c^2\Delta^{\mu\nu}_{\alpha \beta}J_{31}^{(0)-}\dot{u}^{\alpha}\left[\frac{\beta qBb^{\beta \sigma}}{\epsilon+P}V_{\sigma}\right].\nonumber \\
\end{eqnarray}
Similarly for $\mathcal{B}$ we have:
\begin{eqnarray}
\nonumber
\mathcal{B}&=&\Delta^{\mu\nu}_{\alpha \beta}\int_{}^{}dp p^{\alpha}p^{\beta} \frac{\tau_c}{u \cdot p}p^{\rho}\nabla_{\rho}\left[\tau_c \dot{f_0}\right],\\
\nonumber
 &=&-\Delta^{\mu\nu}_{\alpha \beta}\int_{}^{}dp f_0\tilde{f}_0 p^{\alpha}p^{\beta} \frac{\tau_c}{u \cdot p}p^{\rho}\nabla_{\rho}\tau_c\left[ \beta p^{\gamma}\dot{u}_{\gamma}+(u \cdot p)\dot{\beta}-\dot{\alpha}\right].
\end{eqnarray}
Using the thermodynamics integral discussed in Appendix~\ref{appendix1} we get:
\begin{eqnarray}
\label{eq:b}
\nonumber
 \mathcal{B}&=&-2\tau_c^2\left[\left(J_{31}^{(0)+}+J_{42}^{(1)+}\right)\dot{\beta}-\left(J_{31}^{(1)-}+J_{42}^{(2)-}\right)\dot{\alpha}\right]\sigma^{\mu\nu}-2\tau_c^2\nabla^{\langle\mu}\left(\dot{u}^{\nu\rangle}\beta J_{42}^{(1)+}\right), \\
    &=&-2\tau_c^2\left[\left(J_{31}^{(0)+}+J_{42}^{(1)+}\right)\mathcal{X}-\left(J_{31}^{(1)-}+J_{42}^{(2)-}\right)\mathcal{Y}\right]\theta\sigma^{\mu\nu}
 -2\tau_c^2\nabla^{\langle\mu}\left(\dot{u}^{\nu\rangle}\beta J_{42}^{(1)+}\right),
\end{eqnarray}
where in the last line we have used the expression for $\dot{\alpha}$ and $\dot{\beta}$ given in
Eq.~\eqref{eq:alphadot} and Eq.~\eqref{eq:betadot}. The $\mathcal{X}$ and $\mathcal{Y}$ are same as
Eq.~\eqref{lm}.
Finally, for  $\mathcal{C}$ we have
\begin{eqnarray}
\nonumber
\mathcal{C}&=&\Delta^{\mu\nu}_{\alpha \beta}\int_{}^{}dp p^{\alpha}p^{\beta} \frac{\tau_c}{u \cdot p}p^{\rho}\nabla_{\rho} \left[\frac{\tau_c}{u \cdot p}p^{\sigma}\nabla_{\sigma} f_0\right],\\
\nonumber
 &=&-\Delta^{\mu\nu}_{\alpha \beta}\int_{}^{}dp f_0\tilde{f}_0 p^{\alpha}p^{\beta} \frac{\tau_c}{u \cdot p}p^{\rho}\nabla_{\rho} \left[\frac{\tau_c}{u \cdot p}p^{\sigma}\left(\beta p^{\gamma}\nabla_{\sigma}u_{\gamma}+\left(u \cdot p\right)\nabla_{\sigma}\beta-\nabla_{\sigma}\alpha\right)\right].
\end{eqnarray}
Like the previous cases we use the thermodynamic integrals given in Appendix~\ref{appendix1} along with Eq.~\eqref{eq:udot} and Eq.~\eqref{eq:nablaanduexpansion} to rewrite the above expression:
\begin{eqnarray}
\label{eq:c}
\nonumber
  \mathcal{C}&=&2 \nabla^{\langle\mu}\left(\dot{u}^{\nu\rangle}\beta \tau_c^2 J_{42}^{(1)+}\right)+2 \nabla^{\langle\mu}\left[\nabla^{\nu\rangle}\alpha \tau_c^2 \left(J_{42}^{(2)-}-\frac{n_f}{\epsilon+P}J_{42}^{(1)-}\right)\right]
  -4\beta \tau_c^2 \left(2J_{63}^{(3)+}+J_{42}^{(1)+}\right)\sigma^{\langle\mu}_{\rho}\sigma^{\nu\rangle\rho}-\frac{20}{3} \beta \tau_c^2 J_{42}^{(1)+}\theta \sigma^{\mu\nu} \\
  & &
  - \frac{28}{3} \beta \tau_c^2 J_{63}^{(3)+} \theta \sigma^{\mu\nu}-4 \beta \tau_c^2 \left(J_{42}^{(1)+}+2J_{63}^{(3)+}\right) \sigma^{\langle\mu \rho}\omega^{\nu\rangle}_{\rho}
  +2\tau_c^2 \nabla^{\langle\mu}\left[J_{42}^{(1)-}\left(\frac{\beta qBb^{\nu\rangle \gamma}V_\gamma}{\epsilon+P}\right)\right],
\end{eqnarray}
Now let us evaluate the second integral $\mathcal{I}_{2}$:
\begin{eqnarray}
\mathcal{I}_{2}&=&-\Delta^{\mu\nu}_{\alpha\beta}\int_{}^{}dp p^{\alpha}p^{\beta} \left(\left(\frac{\tau_c}{u \cdot p}\right)^2qBb^{\gamma\eta}p_{\eta}\frac{\partial}{\partial p^{\gamma}}\left[p^{\sigma}\partial_{\sigma}f_0\right]\right),\nonumber \\
&=&\Delta^{\mu\nu}_{\alpha\beta}\int_{}^{}dpf_0\tilde{f_0} p^{\alpha}p^{\beta} \left(\left(\frac{\tau_c}{u \cdot p}\right)^2qBb^{\gamma\eta}p_{\eta}\left(\left(\beta p^{\rho}\partial_{\sigma} u_{\rho}+(u \cdot p)\partial_{\sigma} \beta-\partial_{\sigma} \alpha\right)\Delta^{\sigma}_{\gamma}\right)+\beta \partial_{\sigma} u_{\rho} p^{\sigma}\Delta^{\rho}_{\gamma} \right),\nonumber\\
&=&2 \tau_c^2 qB b^{\gamma\eta} \beta J_{42}^{(2)-} \left( \Delta^{\mu\nu}_{\eta\beta}g^{\beta \rho}+ \Delta^{\mu\nu}_{\alpha \eta}g^{\alpha \rho}  \right )\sigma_{\gamma\rho},
\label{eq:I2exp}
\end{eqnarray}
where we have used $\frac{\partial}{\partial p^{\gamma}}p^{\sigma} \partial_{\sigma} f_0 =\partial_{\sigma}f_0 \Delta^{\sigma}_{\gamma} +p^{\sigma}\frac{\partial}{\partial p^{\gamma}} \partial_{\sigma} f_0  $ and the expression for $\partial_{\sigma}f_0$
to arrive at the final expression.

Now using Eqs.~\eqref{eq:a}-\eqref{eq:I2exp} we get the final expression:
\begin{eqnarray}
\label{abcIIshear}
\nonumber
 \pi^{\mu\nu}_{(2)}&=&-\tau_c \dot{\pi}^{\langle\mu\nu\rangle}-2\tau_c^2\dot{u}^{\langle\mu}\nabla^{\nu\rangle}\alpha\left(\frac{n_f}{\epsilon+P}J_{31}^{(0)-}-J_{31}^{(1)-}\right)
  +\tau_c^2\Delta^{\mu\nu}_{\alpha \beta}J_{31}^{(0)-}\dot{u}^{\beta}\left[\frac{\beta qBb^{\alpha \sigma}}{\epsilon+P}V_{\sigma}\right]+\tau_c^2\Delta^{\mu\nu}_{\alpha \beta}J_{31}^{(0)-}\dot{u}^{\alpha}\left[\frac{\beta qBb^{\beta \sigma}}{\epsilon+P}V_{\sigma}\right]\\
 \nonumber
  &&-2\tau_c^2\left[\left(J_{31}^{(0)+}+J_{42}^{(1)+}\right)\mathcal{X}-\left(J_{31}^{(1)-}+J_{42}^{(2)-}\right)\mathcal{Y}\right]\theta\sigma^{\mu\nu}
  -2\tau_c^2\nabla^{\langle\mu}\left(\dot{u}^{\nu\rangle}\beta J_{42}^{(1)+}\right)+2 \nabla^{\langle\mu}\left(\dot{u}^{\nu\rangle}\beta \tau_c^2 J_{42}^{(1)+}\right)\\
  \nonumber
  &&+2 \nabla^{\langle\mu}\left[\nabla^{\nu\rangle}\alpha \tau_c^2 \left(J_{42}^{(2)-}-\frac{n_f}{\epsilon+P}J_{42}^{(1)-}\right)\right] -\frac{20}{3} \beta \tau_c^2 J_{42}^{(1)+}\theta \sigma^{\mu\nu}
  -4\beta \tau_c^2 \left(2J_{63}^{(3)+}+J_{42}^{(1)+}\right)\sigma^{\langle\mu}_{\rho}\sigma^{\nu\rangle\rho}- \frac{28}{3} \beta \tau_c^2 J_{63}^{(3)+} \theta \sigma^{\mu\nu} \\
  &&-4 \beta \tau_c^2 \left(J_{42}^{(1)+}+2J_{63}^{(3)+}\right) \sigma^{\langle\mu \rho}\omega^{\nu\rangle}_{\rho}+2\tau_c^2 \nabla^{\langle\alpha}\left[J_{42}^{(1)-}\left(\frac{\beta qBb^{\beta\rangle \gamma}V_\gamma}{\epsilon+P}\right)\right]
  +2 \tau_c^2 qB b^{\gamma\eta} \beta J_{42}^{(2)-} \left( \Delta^{\mu\nu}_{\eta\beta}g^{\beta \rho}+ \Delta^{\mu\nu}_{\alpha \eta}g^{\alpha \rho}  \right)\sigma_{\gamma\rho}.\nonumber\\
\end{eqnarray}
Here we kept terms only upto second-order in gradients.
\subsection{Bulk stress}
\label{AppSec:bulk}
Let us now consider the bulk viscous case. From Eq.~\eqref{eq:bulk2} we get:
\begin{eqnarray}
\nonumber
\Pi_{(2)}&=&-\frac{1}{3}\Delta_{\alpha\beta}\int_{}^{}dp p^{\alpha}p^{\beta}\left(\frac{\tau_c}{u \cdot p}p^{\mu}\partial_{\mu}\left[\frac{\tau_c}{u \cdot p}p^{\rho}\partial_{\rho}f_0\right]
  +\frac{\tau_c}{u \cdot p}qF^{\mu\nu}p_{\nu}\frac{\partial}{\partial p^{\mu}}\left[\frac{\tau_c}{u \cdot p}p^{\rho}\partial_{\rho}f_0\right]\right), \\
  &=& \mathcal{I}_{1}+\mathcal{I}_{2},
 \end{eqnarray}
where
\begin{eqnarray}
\nonumber
\mathcal{I}_{1}&=&-\frac{\Delta_{\alpha\beta}}{3}\int_{}^{}dp p^{\alpha}p^{\beta}\frac{\tau_c}{u \cdot p}p^{\mu}\partial_{\mu}\left[\frac{\tau_c}{u \cdot p}p^{\rho}\partial_{\rho}f_0\right],\\
\nonumber
\mathcal{I}_{2}&=&-\frac{\Delta_{\alpha\beta}}{3}\int_{}^{}dp p^{\alpha}p^{\beta}\frac{\tau_c}{u \cdot p}qF^{\mu\nu}p_{\nu}\frac{\partial}{\partial p^{\mu}}\left[\frac{\tau_c}{u \cdot p}p^{\rho}\partial_{\rho}f_0\right].
\end{eqnarray}
Note that for our case $F^{\mu\nu}=-Bb^{\mu\nu}$. Let us first evaluate $\mathcal{I}_{1}$ by
breaking it into three parts $\mathcal{I}_{1}=\mathcal{A}+\mathcal{B}+\mathcal{C}$
where
\begin{eqnarray}
\nonumber
\mathcal{A}&=&-\frac{\Delta_{\alpha\beta}}{3}\int_{}^{}dp p^{\alpha}p^{\beta}\tau_c D\left[\frac{\tau_c}{u \cdot p}p^{\rho}\partial_{\rho} f_0\right],\\
\nonumber
\mathcal{B}&=&-\frac{\Delta_{\alpha\beta}}{3}\int_{}^{}dp p^{\alpha}p^{\beta}\frac{\tau_c}{u \cdot p}p^{\mu}\nabla_{\mu}\left(\tau_c \dot{f_0}\right),\\
\nonumber
\mathcal{C}&=&-\frac{\Delta_{\alpha\beta}}{3}\int_{}^{}dp p^{\alpha}p^{\beta}\frac{\tau_c}{u \cdot p}p^{\mu}\nabla_{\mu}\left(\frac{\tau_c p^{\rho}}{u \cdot p}\nabla_{\rho}f_0\right).
\end{eqnarray}
We evaluate each of the above integrals one-by-one:
\begin{eqnarray}
\nonumber
\mathcal{A}&=&-\frac{\Delta_{\alpha\beta}}{3}\int_{}^{}dp p^{\alpha}p^{\beta}\tau_c D\left[\frac{\tau_c}{u \cdot p}p^{\rho}\partial_{\rho} f_0\right],\\
\nonumber
&=&\frac{\Delta_{\alpha\beta}}{3}\int_{}^{}dp f_0 \tilde{f}_0 p^{\alpha}p^{\beta}\tau_c D\left[\frac{\tau_c}{u \cdot p}p^{\rho}\left(\beta p^{\gamma}\partial_{\rho}u_{\gamma}+\left(u \cdot p\right)\partial_{\rho}\beta-\partial_{\rho}\alpha\right)\right], \\
\label{eq:abulk}
  &=&-\tau_c \dot{\Pi}+\frac{2\tau_c^2}{3}J_{31}^{(0)-}\frac{n_f}{\epsilon+P}\nabla^{\alpha}\alpha \dot{u}_{\alpha}-\frac{2\tau_c^2}{3}J_{21}^{(0)-}\nabla^{\alpha}\alpha \dot{u}_{\alpha}
  -\frac{2\tau_c^2\beta}{3(\epsilon+P)}J_{31}^{(0)-}\dot{u}_{\alpha}qBb^{\alpha\beta}V_{\beta}
\end{eqnarray}
We have used the thermodynamic integrals given in Appendix.\ref{appendix1} along with Eq.~\eqref{eq:Iexpansion} and Eq.~\eqref{eq:udot} to arrive at the final expression Eq.~\eqref{eq:abulk}.
Now let us evaluate $\mathcal{B}$ with the help of thermodynamic integrals and its properties given in Appendix~\ref{appendix1}:
\begin{eqnarray}
\nonumber
\mathcal{B}&=&-\frac{\Delta_{\alpha\beta}}{3}\int_{}^{}dp p^{\alpha}p^{\beta}\frac{\tau_c}{u \cdot p}p^{\mu}\nabla_{\mu}(\tau_c \dot{f_0}), \\
\label{eq:bbulk}
&=&\frac{5\tau_c^2}{3}\nabla_{\mu}\left(\beta J_{42}^{(1)+}\dot{u}^{\mu}\right)+\frac{5\tau_c^2}{3}\theta\left[\left(J_{31}^{(0)+}+J_{42}^{(1)+}\right)\dot{\beta}-\left(J_{31}^{(1)-}+J_{42}^{(2)-}\right)\dot{\alpha}\right].
\end{eqnarray}
Similarly for $\mathcal{C}$ we have
\begin{eqnarray}
\nonumber
\mathcal{C}&=&-\frac{\Delta_{\alpha\beta}}{3}\int_{}^{}dp p^{\alpha}p^{\beta}\frac{\tau_c}{u \cdot p}p^{\mu}\nabla_{\mu}(\frac{\tau_c p^{\rho}}{u \cdot p}\nabla_{\rho}f_0),\\
\nonumber
 &=&\frac{\Delta_{\alpha\beta}}{3}\int_{}^{}dp f_0 \tilde{f}_0 p^{\alpha}p^{\beta}\frac{\tau_c}{u \cdot p}p^{\mu}\nabla_{\mu}\left(\frac{\tau_c p^{\rho}}{u \cdot p}\left(\beta p^{\gamma}\nabla_{\rho}u_{\gamma}+\left(u \cdot p\right)\nabla_{\rho}\beta-\nabla_{\rho}\alpha\right)\right), \\
  \nonumber
 &=&\tau_c^2\frac{5}{9}\beta \left(7J_{63}^{(3)+}+\frac{23}{3}J_{42}^{(1)+}\right)\theta^{2} +\tau_c^2\frac{\beta}{3}\left(7J_{63}^{(3)+}+J_{42}^{(1)+}\right)\sigma^{\mu\nu}\sigma_{\mu\nu}
  +\tau_c^2\frac{5}{3}\nabla_{\mu}\left[\nabla^{\mu}\alpha\left(J_{42}^{(1)-}\frac{n_f}{\epsilon+P}-J_{42}^{(2)-}\right)\right] \\
  \label{eq:cbulk}
  &&+\frac{5\tau_c^2}{3}\nabla_{\mu}\left[-J_{42}^{(1)+}\beta\dot{u}^{\mu}-\frac{J_{42}^{(1)-}\beta qBb^{\mu \nu}V_{\nu}}{\epsilon+P}\right].
\end{eqnarray}
Needless to say, here we kept only terms upto the second-order.
The remaining integral $\mathcal{I}_{2}$ is evaluated in a similar fashion,
\begin{eqnarray}
\nonumber
\mathcal{I}_{2}&=&\frac{\Delta_{\alpha\beta}}{3}\int_{}^{}dp p^{\alpha}p^{\beta}\frac{\tau_c}{u \cdot p}qBb^{\mu\nu}p_{\nu}\frac{\partial}{\partial p^{\mu}}\left[\frac{\tau_c}{u \cdot p}p^{\rho}\partial_{\rho}f_0\right],\\
\nonumber
&=&-\frac{\Delta_{\alpha\beta}}{3}\int_{}^{}dp f_0 \tilde{f}_0 p^{\alpha}p^{\beta}\frac{\tau_c}{u \cdot p}qBb^{\mu\nu}p_{\nu}\frac{\partial}{\partial p^{\mu}}\left[\frac{\tau_c}{u \cdot p}p^{\rho}\left(p^{\gamma}\beta \partial_{\rho}u_{\gamma}+(u \cdot p)\partial_{\rho}\beta-\partial_{\rho}\alpha\right)\right], \\
&=&-\frac{\Delta_{\alpha\beta}}{3}\tau_c^2 qBb^{\mu}_{\nu}\left(\beta J_{(2)+}^{\alpha\beta\nu\gamma}\nabla_{\mu}u_{\gamma}+\beta J_{(2)}^{\alpha\beta\nu\rho}\partial_{\rho}u_{\mu}\right),
\end{eqnarray}
using the expansion given in Eq.~\eqref{eq:Iexpansion} and the anti-symmetric property of $b^{\mu\nu}$ we get:
\begin{equation}
\label{eq:IIbulk}
  \mathcal{I}_{2}=-\frac{qB\tau_c^2}{3}\beta J_{42}^{(2)+}\left(5b^{\mu\gamma }\nabla_{\mu}u_{\gamma}+5b^{\mu \rho}\nabla_{\rho}u_{\mu}\right)=0.
\end{equation}
Finally using Eqs.~\eqref{eq:abulk}-\eqref{eq:cbulk} and  Eq.~\eqref{eq:IIbulk} we have:
\begin{eqnarray}
\label{eq:abcIIbulk}
\nonumber
 \Pi_{(2)}&=&-\tau_c \dot{\Pi} +\frac{2\tau_c^2}{3h}J_{31}^{(0)-}\dot{u}_{\alpha}\nabla^{\alpha}\alpha-\frac{2\tau_{c}^2}{3}J_{21}^{(0)-}\dot{u}_{\alpha}\nabla^{\alpha}\alpha
  -\frac{2\tau_c^2\beta}{3\left(\epsilon+P\right)}J_{31}^{(0)-}\dot{u}_{\alpha}qBb^{\alpha\beta}V_{\beta}+\frac{5\tau_{c}^2 \beta}{9} \left(7J_{63}^{(3)+}+\frac{23}{3}J_{42}^{(1)+}\right)\theta^2
  \\
   \nonumber
   && +\frac{5\tau_c^2}{3}\left[\left(J_{31}^{(0)+}+J_{42}^{(1)+}\right)\dot{\beta}-\left(J_{31}^{(1)-}+J_{42}^{(2)-}\right)\dot{\alpha}\right]\theta
   +\frac{\tau_c^2 \beta}{3}\left(7J_{63}^{(3)+}+J_{42}^{(1)+}\right)\sigma^{\mu\nu}\sigma_{\mu\nu}
  \\
  &&+\frac{5\tau_c^2}{3}\nabla_{\mu}\left[\left( \frac{1}{h}J_{42}^{(1)-}-J_{42}^{(2)-}\right)\nabla^{\mu}\alpha\right]-\frac{5\tau_c^2}{3}\nabla_{\mu}\left[\frac{J_{42}^{(1)-}\beta qBb^{\mu \nu}V_{\nu}}{\epsilon+P}\right].
\end{eqnarray}
Eq.~\eqref{eq:abcIIbulk} is the second-order relaxation equation for the bulk-viscous stress.
\subsection{Diffusion current}
\label{AppSubsec:Diffusion}
In this section we discuss the detail derivation of the diffusion current. From Eq.~\eqref{eq:Diffusion2} we get:
\begin{eqnarray}
\nonumber
V^{\mu}_{(2)}&=&\Delta^{\mu}_{\alpha}\int_{}^{}dp p^{\alpha}\left(\frac{\tau_c}{u \cdot p}p^{\sigma}\partial_{\sigma}\left[\frac{\tau_c}{u \cdot p}p^{\rho}\partial_{\rho}f_0\right]
  +\frac{\tau_c}{u \cdot p}qF^{\sigma\nu}p_{\nu}\frac{\partial}{\partial p^{\sigma}}\left[\frac{\tau_c}{u \cdot p}p^{\rho}\partial_{\rho}f_0\right]\right), \\
  &=& \mathcal{I}_{1} + \mathcal{I}_{2}.
\end{eqnarray}
where
\begin{eqnarray}
\nonumber
\mathcal{I}_{1}&=&\Delta^{\mu}_{\alpha}\int_{}^{}dp p^{\alpha}\frac{\tau_c}{u \cdot p}p^{\sigma}\partial_{\sigma}\left[\frac{\tau_c}{u \cdot p}p^{\rho}\partial_{\rho}f_0\right],\\
\nonumber
\mathcal{I}_{2}&=&\Delta^{\mu}_{\alpha}\int_{}^{}dp p^{\alpha}\frac{\tau_c}{u \cdot p}qF^{\sigma\nu}p_{\nu}\frac{\partial}{\partial p^{\sigma}}\left[\frac{\tau_c}{u \cdot p}p^{\rho}\partial_{\rho}f_0\right].
\end{eqnarray}
Let us first calculate the $\mathcal{I}_{1}$ by breaking it up into three parts as:
$\mathcal{I}_{1}=\mathcal{A}+\mathcal{B}+\mathcal{C}$ where
\begin{eqnarray}
\nonumber
 \mathcal{A}&=&\Delta^{\mu}_{\alpha}\int_{}^{}dp p^{\alpha}\tau_c D\left[\frac{\tau_c}{u \cdot p}p^{\rho}\partial_{\rho} f_0\right],\\
\nonumber
\mathcal{B}&=&\Delta^{\mu}_{\alpha}\int_{}^{}dp p^{\alpha}\frac{\tau_c}{u \cdot p}p^{\sigma}\nabla_{\sigma}\left(\tau_c \dot{f_0}\right),\\
\nonumber
\mathcal{C}&=&\Delta^{\mu}_{\alpha}\int_{}^{}dp p^{\alpha}\frac{\tau_c}{u \cdot p}p^{\sigma}\nabla_{\sigma}\left(\frac{\tau_c p^{\rho}}{u \cdot p}\nabla_{\rho}f_0\right).
\end{eqnarray}
 For $\mathcal{A}$ we get:
\begin{eqnarray}
\nonumber
\mathcal{A}&=&-\Delta^{\mu}_{\alpha}\int_{}^{}dpf_0 \tilde{f}_0 p^{\alpha}\tau_c D\left[\frac{\tau_c}{u \cdot p}p^{\rho} \left(\beta p^{\gamma}\partial_{\rho}u_{\gamma}+\left(u \cdot p\right)\partial_{\rho}\beta-\partial_{\rho}\alpha\right)\right],\\
\nonumber
\label{eq:adiffusion}
&=&-\Delta^{\mu}_{\alpha}D\left[\int_{}^{}dpf_0 \tilde{f}_0 p^{\alpha}\tau_c \frac{\tau_c}{u \cdot p}p^{\rho} \left(\beta p^{\gamma}\partial_{\rho}u_{\gamma}+\left(u \cdot p\right)\partial_{\rho}\beta-\partial_{\rho}\alpha\right)\right], \\
&=&-\tau_c \dot{V}^{\langle\mu\rangle}- \tau_c^2\Delta^{\mu}_{\gamma}D\left[ \frac{n_f qBb^{\gamma \nu}V_{\nu}}{\epsilon+P} \right].
\end{eqnarray}
We have used the thermodynamic integrals and its expansion given in the Appendix~\ref{appendix1},
along with Eqs.~\eqref{eq:alphadot} and~\eqref{eq:betadot} to arrive at the final expression.
Similarly for $\mathcal{B}$ we get:
\begin{eqnarray}
\mathcal{B}&=&\Delta^{\mu}_{\alpha}\int_{}^{}dp p^{\alpha}\frac{\tau_c}{u \cdot p}p^{\sigma}\nabla_{\sigma}\left(\tau_c \dot{f_0}\right),\nonumber\\
\nonumber
&=&\Delta^{\mu}_{\alpha}\nabla_{\sigma}\left(\int_{}^{}dp p^{\alpha}\frac{\tau_c}{u \cdot p}p^{\sigma}\tau_c \dot{f_0}\right)
+\Delta^{\mu}_{\alpha}\nabla_{\sigma}u_{\gamma}\left(\int_{}^{}dp p^{\alpha}p^{\gamma}\frac{\tau_c}{(u \cdot p)^2}p^{\sigma}\tau_c \dot{f_0}\right), \\
\label{eq:bdiff}
&=&-\tau_c^2 \nabla^{\mu}\left(J_{21}^{(0)-}\dot{\beta}-J_{21}^{(1)+}\dot{\alpha}\right)-\tau_c^2 \beta \dot{u}^{\mu}\theta \left(\frac{4}{3}J_{21}^{(0)-}+\frac{5}{3}J_{42}^{(2)-}\right)-\tau_c^2 \beta J_{21}^{(0)-}\dot{u}_{\gamma} \omega^{\gamma\mu}-\tau_c^2\beta \dot{u}_{\gamma}\sigma^{\gamma\mu}\left(J_{21}^{(0)-}+2J_{42}^{(2)-}\right).\nonumber\\
\end{eqnarray}
Lastly for $\mathcal{C}$ we get:
\begin{eqnarray}
\nonumber
\mathcal{C}&=&\Delta^{\mu}_{\alpha}\int_{}^{}dp p^{\alpha}\frac{\tau_c^2}{u \cdot p}p^{\sigma}
\nabla_{\sigma}\left(\frac{p^{\rho}}{u \cdot p}\nabla_{\rho}f_0\right),\\ \nonumber
&=&\Delta^{\mu}_{\alpha}\nabla_{\sigma}\left(\int_{}^{}dp p^{\alpha}\frac{\tau_c^2}
{u \cdot p}p^{\sigma}\frac{p^{\rho}}{u \cdot p}\nabla_{\rho}f_0\right)
+\Delta^{\mu}_{\alpha}\nabla_{\sigma}u_{\gamma}\left(\int_{}^{}dp p^{\alpha}p^{\gamma}
\frac{\tau_c^2}{\left(u \cdot p\right)^2}p^{\sigma}\frac{p^{\rho}}{u \cdot p}\nabla_{\rho}f_0\right).
\end{eqnarray}
Substituting the expression for $\nabla_{\rho}f_0$ and using the usual thermodynamic integrals and their expansion
along with Eq.~\eqref{eq:udot} and Eq.~\eqref{eq:nablaanduexpansion} the above expression takes the following form:
\begin{eqnarray}
\label{eq:cdiff}
 \mathcal{C}&=&-\frac{4\tau_c^2}{3}\left(\frac{J_{21}^{(0)+}n_f}{\epsilon+P}-J_{21}^{(1)+}\right)\left(\nabla^{\mu}\alpha\right) \theta +\frac{4\tau_c^2}{3}J_{21}^{(0)-}\beta\dot{u}^{\mu}\theta-\tau_c^2\left(\frac{J_{21}^{(0)+}n_f}{\epsilon+P}-J_{21}^{(1)+}\right)\left(\nabla^{\gamma}\alpha\right) \sigma^{\mu}_{\gamma} +\tau_c^2J_{21}^{(0)-}\beta\dot{u}^{\gamma}\sigma^{\mu}_{\gamma}\nonumber\\
 &&-\tau_c^2\left(\frac{J_{21}^{(0)+}n_f}{\epsilon+P}-J_{21}^{(1)+}\right)\left(\nabla^{\gamma}\alpha\right) \omega^{\mu}_{\gamma} +\tau_c^2J_{21}^{(0)-}\beta\dot{u}^{\gamma}\omega^{\mu}_{\gamma}-2\tau_c^2\left(\frac{J_{42}^{(2)+}n_f}{\epsilon+P}-J_{42}^{(3)+}\right)\left(\nabla^{\gamma}\alpha\right) \sigma^{\mu}_{\gamma} +2\tau_c^2J_{42}^{(2)-}\beta\dot{u}^{\gamma}\sigma^{\mu}_{\gamma}\nonumber\\
 &&-\frac{5\tau_c^2}{3}\left(\frac{J_{42}^{(2)+}n_f}{\epsilon+P}-J_{42}^{(3)+}\right)\left(\nabla^{\mu}\alpha\right) \theta +\frac{5\tau_c^2}{3}J_{42}^{(2)-}\beta\dot{u}^{\mu}\theta-2\tau_c^2\Delta^{\mu}_{\rho} \nabla_{\gamma}\left(\beta J_{42}^{(2)-}\sigma^{\rho\gamma}\right)-\frac{5\tau_c^2}{3}\nabla^{\mu}\left[\beta J_{42}^{(2)-}\theta\right]\nonumber\\
 &&+\frac{4\tau_c}{3}J_{21}^{(0)+} \theta\left[\frac{\beta qBb^{\mu\nu}V_{\nu}}{\epsilon+P}\right]+\tau_c J_{21}^{(0)+} \sigma^{\mu}_{\gamma}\left[\frac{\beta qBb^{\gamma\nu}V_{\nu}}{\epsilon+P}\right] +2\tau_c J_{42}^{(2)+} \sigma^{\mu}_{\gamma} \left[\frac{\beta qBb^{\gamma\nu}V_{\nu}}{\epsilon+P}\right]+\frac{5\tau_c}{3}J_{42}^{(2)+}\theta \left[\frac{\beta qBb^{\mu\nu}V_{\nu}}{\epsilon+P}\right] \nonumber\\
 &&+\tau_c J_{21}^{(0)+} \omega^{\mu}_{\gamma}\left[\frac{\beta qBb^{\gamma\nu}V_{\nu}}{\epsilon+P}\right]
\end{eqnarray}
Now let us calculate the integral $\mathcal{I}_{2}$:
\begin{eqnarray}
\nonumber
\mathcal{I}_{2}&=&-\Delta^{\mu}_{\alpha}\int_{}^{}dp f_0 \tilde{f}_0 p^{\alpha}\frac{\tau_c}{u \cdot p}qF^{\sigma\nu}p_{\nu}\frac{\partial}{\partial p^{\sigma}}\left[\frac{\tau_c}{u \cdot p}p^{\rho}\left(\beta p^{\gamma}\partial_{\rho}u_{\gamma}+(u \cdot p)\partial_{\rho}\beta-\partial_{\rho}\alpha\right)\right] ,\\
\label{eq:IIdiffusion}
  &=&\tau_c^2 qB\left[\frac{n_f J_{21}^{(1)-}b^{\gamma\mu}}{\epsilon+P}\nabla_{\gamma}\alpha-J_{21}^{(2)-}b^{\gamma\mu}\nabla_{\gamma}\alpha
  -\frac{\beta J_{21}^{(1)+}b^{\gamma\mu}\Delta^{\sigma}_{\gamma}\partial^k\pi_{k\sigma}}{\epsilon+P}-\frac{\beta J_{21}^{(1)+}b^{\gamma\mu}\Pi\dot{u}_{\gamma}}{\epsilon+P}
  +\frac{\beta J_{21}^{(1)+}b^{\gamma\mu}\nabla_{\gamma} \Pi}{\epsilon+P}\right].
\end{eqnarray}

Adding Eqs.~\eqref{eq:adiffusion}-\eqref{eq:IIdiffusion} together we get
the final expression for the diffusion current (which after simplification  becomes Eq.\eqref{eq:diffusionEvolution2} )  :
\begin{eqnarray}
\nonumber
V^{\mu}_{(2)}&=&-\tau_c \dot{V}^{\langle\mu\rangle}- \tau_c^2\Delta^{\mu}_{\gamma}D\left[ \frac{n_f qBb^{\gamma \nu}V_{\nu}}{\epsilon+P} \right] -\tau_c^2 \nabla^{\mu}\left(J_{21}^{(0)-}\dot{\beta}-J_{21}^{(1)+}\dot{\alpha}\right)
-\tau_c^2 \beta \dot{u}^{\mu}\theta \left(\frac{4}{3}J_{21}^{(0)-}+\frac{5}{3}J_{42}^{(2)-}\right)\\
\nonumber
& & -\tau_c^2 \beta J_{21}^{(0)-}\dot{u}_{\gamma} \omega^{\gamma\mu}
-\tau_c^2\beta \dot{u}_{\gamma}\sigma^{\gamma\mu}\left(J_{21}^{(0)-}+2J_{42}^{(2)-}\right)-\frac{4\tau_c^2}{3}\left(\frac{J_{21}^{(0)+}n_f}{\epsilon+P}-J_{21}^{(1)+}\right)
\left(\nabla^{\mu}\alpha\right) \theta   +\frac{4\tau_c^2}{3}J_{21}^{(0)-}\beta\dot{u}^{\mu}\theta \\
& & -\tau_c^2\left(\frac{J_{21}^{(0)+}n_f}{\epsilon+P}-J_{21}^{(1)+}\right)
\left(\nabla^{\gamma}\alpha\right) \sigma^{\mu}_{\gamma} +\tau_c^2J_{21}^{(0)-}
\beta\dot{u}^{\gamma}\sigma^{\mu}_{\gamma}-\tau_c^2\left(\frac{J_{21}^{(0)+}n_f}{\epsilon+P}-J_{21}^{(1)+}\right)
 \left(\nabla^{\gamma}\alpha\right) \omega^{\mu}_{\gamma}+ \tau_c^2J_{21}^{(0)-}\beta\dot{u}^{\gamma}\omega^{\mu}_{\gamma} \nonumber\\
 &&-2\tau_c^2\left(\frac{J_{42}^{(2)+}n_f}{\epsilon+P}-J_{42}^{(3)+}\right)
 \left(\nabla^{\gamma}\alpha\right) \sigma^{\mu}_{\gamma} 
 +2\tau_c^2J_{42}^{(2)-}\beta\dot{u}^{\gamma}\sigma^{\mu}_{\gamma}-\frac{5\tau_c^2}{3}\left(\frac{J_{42}^{(2)+}n_f}{\epsilon+P}-J_{42}^{(3)+}\right)
 \left(\nabla^{\mu}\alpha\right) \theta  +\frac{5\tau_c^2}{3}J_{42}^{(2)-}\beta\dot{u}^{\mu}\theta\nonumber\\
 && -2\tau_c^2\Delta^{\mu}_{\rho} \nabla_{\gamma}\left(\beta J_{42}^{(2)-}\sigma^{\rho\gamma}\right)
 -\frac{5\tau_c^2}{3}\nabla^{\mu}\left[\beta J_{42}^{(2)-}\theta\right]+\frac{4\tau_c}{3}J_{21}^{(0)+} 
 \theta\left[\frac{\beta qBb^{\mu\nu}V_{\nu}}{\epsilon+P}\right]  +\tau_c J_{21}^{(0)+} \sigma^{\mu}_{\gamma}
 \left[\frac{\beta qBb^{\gamma\nu}V_{\nu}}{\epsilon+P}\right] \nonumber\\
 &&+2\tau_c J_{42}^{(2)+} \sigma^{\mu}_{\gamma} 
 \left[\frac{\beta qBb^{\gamma\nu}V_{\nu}}{\epsilon+P}\right]
 +\frac{5\tau_c}{3}J_{42}^{(2)+}\theta \left[\frac{\beta qBb^{\mu\nu}V_{\nu}}{\epsilon+P}\right]
   +\tau_c J_{21}^{(0)+} \omega^{\mu}_{\gamma}\left[\frac{\beta qBb^{\gamma\nu}V_{\nu}}{\epsilon+P}
 \right] \nonumber \\
 &&+ \tau_c^2 qB\left[\frac{n_f J_{21}^{(1)-}b^{\gamma\mu}}{\epsilon+P}\nabla_{\gamma}\alpha-
 J_{21}^{(2)-}b^{\gamma\mu}\nabla_{\gamma}\alpha
  -\frac{\beta J_{21}^{(1)+}b^{\gamma\mu}\Delta^{\sigma}_{\gamma}\partial^k\pi_{k\sigma}}
  {\epsilon+P}-\frac{\beta J_{21}^{(1)+}b^{\gamma\mu}\Pi\dot{u}_{\gamma}}{\epsilon+P}
  +\frac{\beta J_{21}^{(1)+}b^{\gamma\mu}\nabla_{\gamma} \Pi}{\epsilon+P}\right].
\end{eqnarray}

\newpage
\section{General expressions for transport coefficients in terms of thermodynamic integrals}{\label{appendix3}}
\begin{table*}[h]
\begin{center}
\begin{tabular}{|p{1.5cm}|p{11.0cm}|}
 \hline
 $\tau_{\pi\pi}$&$\frac{2\beta}{\beta_{\pi}}\left(2J_{63}^{(3)+}+J_{42}^{(1)+}\right)$\\
  \hline
  $\delta_{\pi\pi}$&$\frac{\beta}{3 \beta_{\pi}}\left(7J_{63}^{(3)+}+5J_{42}^{(1)+}
  \right)$\\
  \hline
  $\lambda_{\pi\Pi}$&$\frac{2}{\beta_{\Pi}}\left[\left(J_{31}^{(0)+}+J_{42}^{(1)+}\right)\mathcal{X}-\left(J_{31}^{(1)-}+J_{42}^{(2)-}\right)\mathcal{Y} +\frac{\beta}{3}\left(7J_{63}^{(3)+}+5J_{42}^{(1)+}\right)\right]$\\
  \hline
  $l_{\pi V}$&$\frac{2}{\beta_V }\left(J_{42}^{(2)-}-\frac{n_f}{\epsilon+P}J_{42}^{(1)-}\right)$\\
  \hline
  $\delta_{\pi B}$&${2 J_{42}^{(2)-}}/{J_{42}^{(1)+}}$\\
  \hline
  $ \delta_{\pi VB}$&${2\beta J_{42}^{(1)-}}/{\left(\epsilon+P\right)}$\\
\hline
 \end{tabular}
\\[13pt]
\caption{Transport coefficients appearing in shear-stress equation Eq.~\eqref{eq:shear_evolution2}.}
\label{table:ShearJmn}
\end{center}
\end{table*}
\begin{table*}[h]
\begin{center}
\begin{tabular}{|p{1.5cm}|p{9.0cm}|}
 \hline
$\delta_{\Pi\Pi}$&$\frac{5}{3\beta_{\Pi}}\big[(J_{31}^{(0)+}+J_{42}^{(1)+})\mathcal{X}-(J_{31}^{(1)-}+J_{42}^{(2)-})\mathcal{Y} +\frac{\beta}{3}(7J_{63}^{(3)+}+\frac{23}{5} J_{42}^{(1)+})\big]$ \\
\hline
$\lambda_{\Pi\pi}$&$\frac{\beta}{3\beta_{\pi}}\left(7J_{63}^{(3)+}+J_{42}^{(1)+}\right)$ \\
\hline
$l_{\Pi V}$&$\frac{5}{3\beta_V}\left(J_{42}^{(2)-}-\frac{n_f}{\epsilon+P}J_{42}^{(1)-}\right)$\\\hline
$\delta_{\Pi V B}$&$\frac{5J_{42}^{(1)-}\beta}{3\left(\epsilon+P\right)}$\\
\hline
 \end{tabular}
\\[13pt]
\caption{Transport coefficients appearing in bulk equation Eq.~\eqref{eq: bulk_evolution2}.}
\label{table:BulkJmn}
\end{center}
\end{table*}
\begin{table*}[h]
\begin{center}
\begin{tabular}{|p{1.5cm}|p{8.0cm}|}
 \hline
$\lambda_{VV}$&$1+\frac{2}{\beta_V}\left(\frac{n_f}{\epsilon+P}J_{42}^{(2)+}-J_{42}^{(3)+}\right)$\\
\hline
$ \delta_{VV}$&$\frac{4}{3}+\frac{5}{3\beta_V}\left(\frac{n_f J_{42}^{(2)+}}{\epsilon+P}-J_{42}^{(3)+}\right)$\\
\hline
$l_{V\pi}$&$-\frac {\beta}{\beta_{\pi}} J_{42}^{(2)-}$\\
\hline
$l_{V\Pi}$&$-\frac{1}{\beta_{\Pi}}\left(\mathcal{X} J_{21}^{(0)-}-\mathcal{Y} J_{21}^{(1)+} +\frac{5\beta}{3} J_{42}^{(1)-}\right)$\\
\hline
$\delta_{VB}$&$-\left(\frac{n_f J_{21}^{(1)-}}{\epsilon+P}-J_{21}^{(2)-}\right)/\beta_{V}$\\
\hline
$l_{V\pi B}$&$-{\beta J_{21}^{(1)+}}/{\left(\epsilon+P\right)}$\\
\hline
$\tau_{V \Pi B}$&$-{\beta J_{21}^{(1)+}}/{\left(\epsilon+P\right)}$\\
\hline
$l_{V \Pi B}$&$-{\beta J_{21}^{(1)+}}/{\left(\epsilon+P\right)}$\\
\hline
$\lambda_{VVB}$&$-\frac{\beta}{\epsilon+P}\left(J_{21}^{(0)+}+2J_{42}^{(2)+}\right)$\\
\hline
$\mathbf{\rho}_{VVB}$&$-{\beta J_{21}^{(0)+}}/{\left(\epsilon+P\right)}$\\
\hline
$\tau_{VVB}$&${n_f}/\left({\epsilon+P}\right)$\\
\hline
 \end{tabular}
\\[13pt]
\caption{Transport coefficients appearing in diffusion equation Eq.~\eqref{eq:diffusionEvolution2}.}
\label{table:DiffusionJmn}
\end{center}
\end{table*}

\bibliography{Ref}

\end{document}